\documentclass[useAMS,usenatbib]{mn2e}
\usepackage[dvips]{graphicx}
\usepackage{amsmath}
\usepackage{hyperref}
\usepackage{xcolor}
\newcommand{\bz}{$\langle B_z \rangle$}
\newcommand{\nz}{$\langle N_z \rangle$}
\newcommand{\vsini}{$v \sin i$}
\newcommand{\kms}{km\,s$^{-1}$}

\newcommand{\rsun}{R$_\odot$}
\newcommand{\msun}{M$_\odot$}

\newcommand{\teff}{$T_{\rm eff}$}
\newcommand{\ra}{$R_{\rm A}$}
\newcommand{\rk}{$R_{\rm K}$}
\newcommand{\rark}{$\log{(R_{\rm A}/R_{\rm K})}$}

\newcommand{\mnras}{$MNRAS$}
\newcommand{\apj}{$ApJ$}
\newcommand{\apjl}{$ApJL$}
\newcommand{\aj}{$AJ$}
\newcommand{\aap}{A\&A}

\newcommand{\aaps}{A\&AS}

\newcommand{\apjs}{ApJS}
\newcommand{\araa}{ARAA}

\newcommand{\nat}{Nature}

\newcommand{\physscr}{Phys.\ Scr.}

\title[W\,601: a magnetospheric hot PMS binary]{NGC\,6611\,601: A hot pre-main sequence spectroscopic binary containing a centrifugal magnetosphere host star}
\author[M.\ E.\ Shultz]
{M.\ E.\ Shultz$^{1}$\thanks{E-mail: mshultz@udel.edu},
E.\ Alecian$^2$,
V.\ Petit$^1$,
S.\ Bagnulo$^3$,
T. B\"ohm$^{4,5}$,
C.\ P.\ Folsom$^{5,6}$, 
\newauthor{
G.\ A.\ Wade$^6$, 
and the MiMeS Collaboration} \\
$^1$Department of Physics and Astronomy, University of Delaware, 217 Sharp Lab, Newark, Delaware, 19716, USA\\
$^2$Universit\'e Grenoble Alpes, IPAG, F-38000 Grenoble, France\\
$^4$Universit\'e de Toulouse; UPS-OMP; IRAP; Toulouse, France\\
$^5$CNRS; IRAP; 14, avenue Edouard Belin, 31400 Toulouse, France\\
$^6$Department of Physics and Space Science, Royal Military College of Canada, Kingston, Ontario K7K 7B4, Canada\\
}

\begin{document}

\date{}

\pagerange{\pageref{firstpage}--\pageref{lastpage}} \pubyear{2018}

\maketitle

\label{firstpage}

\begin{abstract}
W\,601 (NGC\,6611\,601) is one of the handful of known magnetic Herbig Ae/Be stars. We report the analysis of a large dataset of high-resolution spectropolarimetry. The star is a previously unreported spectroscopic binary, consisting of 2 B2 stars with a mass ratio of 1.8, masses of 12 \msun~and 6.2 \msun, in an eccentric 110-day orbit. The magnetic field belongs to the secondary, W\,601\,B. The H$\alpha$ emission is consistent with an origin in W\,601\,B's centrifugal magnetosphere; the star is therefore not a classical Herbig Be star in the sense that its emission is not formed in an accretion disk. However, the low value of $\log{g} = 3.8$ determined via spectroscopic analysis, and the star's membership in the young NGC\,6611 cluster, are most consistent with it being on the pre-main sequence. The rotational period inferred from the variability of the H$\alpha$ line and the longitudinal magnetic field \bz~is 1.13~d. Modelling of Stokes $V$ and \bz~indicates a surface dipolar magnetic field $B_{\rm d}$ between 6 and $11$~kG. With its strong emission, rapid rotation, and strong surface magnetic field, W\,601\,B is likely a precursor to H$\alpha$-bright magnetic B-type stars such as $\sigma$ Ori E. By contrast, the primary is an apparently non-magnetic ($B_{\rm d} < 300$~G) pre-main sequence early B-type star. In accordance with expectations from magnetic braking, the non-magnetic primary is apparently more rapidly rotating than the magnetic star.
\end{abstract}


\section{Introduction}

Approximately 10\% of main sequence and pre-main sequence stars with radiative envelopes host strong ($\sim$kG scale), organized (predominantly dipolar) magnetic fields \citep{2017MNRAS.465.2432G,2019MNRAS.483.2300S}. Since radiative envelopes cannot support a contemporaneous convective dynamo, it is believed that the magnetic fields of early type stars are fossils -- remnants from a previous stage in the stars' evolution \citep[e.g.][]{2004Natur.431..819B,2015IAUS..305...61N}. Consistent with the fossil field hypothesis is that, unlike dynamo magnetic fields, hot star magnetic fields are apparently stable over a time-span of at least decades \citep[e.g.][]{2018MNRAS.475.5144S}, show no general correlation between magnetic field strength and rotational properties, and exhibit a decline in surface magnetic field strength over evolutionary timescales consistent with conservation or slow decay of magnetic flux \citep[e.g.][]{land2007,land2008,2019MNRAS.483.3127S,2019MNRAS.490..274S}.

The origin of fossil magnetic fields remains obscure. One scenario is that they arise due to short-lived dynamos generated during binary mergers \citep[e.g.][]{2019Natur.574..211S}, however this is difficult to reconcile with the orbital properties of known close magnetic binary systems such as the doubly-magnetic system $\epsilon$ Lupi \citep{2015MNRAS.454L...1S}, the tidally locked system HD\,98088 \citep{2013MNRAS.431.1513F}, the very close binary HD\,156324 \citep[][]{2018MNRAS.475..839S}, or the `identical twins' of HD\,62658 \citep{2019MNRAS.490.4154S}. A competing scenario is that fossil fields may be remnants of convective dynamos operating on the pre-main sequence (PMS) \citep[e.g.][]{1999stma.book.....M}, in which case the 10\% incidence ratio might perhaps be explained by a PMS dynamo bistability mechanism similar to that seen in fully convective stars \citep[as discussed by][]{2019MNRAS.490.4154S} combined with rapid rotationally or convectively driven decay of magnetic fields failing to reach critical surface strength \citep{2007A&A...475.1053A,2020ApJ...900..113J}.

\begin{table*}
\centering
\caption[]{Observation log, RV measurements, and \bz~measurements. $S/N$ indicates the maximum signal-to-noise per spectral pixel in the ESPaDOnS spectrum. Average RV uncertainties are 1.8 \kms~for W\,601\,A and 1.5 \kms~for W\,601\,B. DF refers to the magnetic field detection flag (DD: Definite Detection; MD: Marginal Detection; ND: Non-Detection).}
\label{obstab}
\begin{tabular}{l r r r r r r r r}
\hline\hline
Date & ${\rm HJD}-$ & $S/N$ & ${\rm RV_A}$   & ${\rm RV_B}$   & \bz & DF$_V$ & \nz & DF$_N$ \\
     & $-2453000$   &     & (${\rm km/s}$) & (${\rm km/s}$) & (G) &        & (G) &        \\
\hline
10/08/2006 &  957.76004 & 220 & $5$ & $20$ & $184 \pm 286$ &  DD & $-139 \pm 286$ &  ND \\
02/03/2007 & 1162.15478 & 103 & $-7$ & $38$ & $-709 \pm 1015$ &  ND & $-1067 \pm 1015$ &  ND \\
03/03/2007 & 1163.12053 & 192 & $0$ & $31$ & $408 \pm 348$ &  ND & $-644 \pm 348$ &  ND \\
04/03/2007 & 1164.12133 & 226 & $-2$ & $34$ & $-857 \pm 286$ &  ND & $-391 \pm 286$ &  ND \\
05/03/2007 & 1165.13840 & 254 & $-6$ & $41$ & $-805 \pm 247$ &  ND & $-450 \pm 247$ &  ND \\
06/03/2007 & 1166.10187 & 179 & $-6$ & $45$ & $-511 \pm 343$ &  ND & $116 \pm 343$ &  ND \\
06/03/2007 & 1166.14879 & 153 & $-7$ & $47$ & $-458 \pm 617$ &  ND & $213 \pm 617$ &  ND \\
07/03/2007 & 1167.11236 & 182 & $-9$ & $49$ & $-117 \pm 362$ &  MD & $59 \pm 362$ &  ND \\
08/03/2007 & 1168.10114 & 194 & $-6$ & $45$ & $1424 \pm 354$ &  DD & $-508 \pm 354$ &  ND \\
08/03/2007 & 1168.14652 & 204 & $-9$ & $47$ & $921 \pm 332$ &  DD & $606 \pm 332$ &  ND \\
09/03/2007 & 1169.10634 & 190 & $0$ & $38$ & $2014 \pm 349$ &  DD & $-813 \pm 348$ &  ND \\
09/03/2007 & 1169.15224 & 199 & $-1$ & $36$ & $2036 \pm 370$ &  DD & $-246 \pm 369$ &  ND \\
30/06/2008 & 1647.80106 & 203 & $35$ & $-30$ & $2095 \pm 311$ &  DD & $547 \pm 311$ &  ND \\
30/06/2008 & 1647.84933 & 191 & $35$ & $-30$ & $2079 \pm 335$ &  DD & $336 \pm 334$ &  ND \\
30/06/2008 & 1647.89851 & 185 & $33$ & $-28$ & $1784 \pm 354$ &  DD & $31 \pm 354$ &  ND \\
30/06/2008 & 1647.94664 & 175 & $31$ & $-25$ & $1797 \pm 393$ &  DD & $-589 \pm 392$ &  ND \\
30/06/2008 & 1647.99560 & 179 & $28$ & $-19$ & $443 \pm 359$ &  MD & $-283 \pm 359$ &  ND \\
30/06/2008 & 1648.04373 & 179 & $29$ & $-19$ & $1178 \pm 382$ &  DD & $-33 \pm 382$ &  ND \\
01/07/2008 & 1648.80655 & 209 & $31$ & $-30$ & $1446 \pm 311$ &  DD & $200 \pm 310$ &  ND \\
01/07/2008 & 1648.85518 & 205 & $31$ & $-26$ & $1808 \pm 325$ &  DD & $17 \pm 324$ &  ND \\
01/07/2008 & 1648.90492 & 202 & $32$ & $-28$ & $2509 \pm 327$ &  DD & $-499 \pm 326$ &  ND \\
01/07/2008 & 1648.95328 & 205 & $31$ & $-25$ & $2359 \pm 320$ &  DD & $-212 \pm 320$ &  ND \\
01/07/2008 & 1649.00232 & 219 & $33$ & $-28$ & $2053 \pm 291$ &  DD & $6 \pm 290$ &  ND \\
01/07/2008 & 1649.05057 & 213 & $31$ & $-25$ & $1748 \pm 299$ &  DD & $101 \pm 299$ &  ND \\
29/07/2008 & 1676.89812 & 167 & $9$ & $15$ & $370 \pm 412$ &  ND & $-430 \pm 412$ &  ND \\
21/06/2012 & 3099.93880 & 272 & $24$ & $-8$ & $2088 \pm 308$ &  DD & $-170 \pm 307$ &  ND \\
13/08/2013 & 3517.80010 & 264 & $34$ & $-27$ & $1256 \pm 329$ &  DD & $480 \pm 329$ &  ND \\
\hline\hline
\end{tabular}
\end{table*}

This makes examination of magnetic hot stars on the PMS a key arena for determining the origin of fossil magnetic fields. Only eight PMS early-type stars with magnetic field detections confirmed with high-resolution spectropolarimetry are known \citep{2005A&A...442L..31W, 2008MNRAS.387L..23P, 2007A&A...462..293C, 2008MNRAS.385..391A, 2008A&A...481L..99A, 2009A&A...502..283H, 2013MNRAS.429.1001A, 2013AN....334.1093H, 2015MNRAS.449L.118H, 2015A&A...584A..15J}. \cite{2013MNRAS.429.1001A,2013MNRAS.429.1027A} showed that fossil magnetic fields have the same incidence amongst Herbig Ae/Be stars as amongst the MS population, and tend to be more slowly rotating than non-magnetic Herbig stars, which is consistent with the magnetic stars being subject to magnetic braking. A spectropolarimetric survey of intermediate mass T Tauri stars by \cite{2019A&A...622A..72V} demonstrated that magnetic fields are ubiquitous amongst stars with convective envelopes, but that the incidence declines to around 10\% immediately upon crossing the boundary on the Hertzsprung-Russell diagram between convective and radiative envelopes. 

Since only a few magnetic PMS hot stars are known, a full magnetic and rotational characterization of the individual members of this population is an important step in understanding their properties. In this paper, we examine the PMS B2 star NGC\,6611\,601 (also known as NGC\,6611-019, ALS\,9522, BD\,$-13^\circ~4937$, and hereafter referred to as W\,601). This star was classified as a Herbig Ae/Be star on the basis of its `P-Cygni like' H$\alpha$ emisison \citep{2008A&A...489..459M} and its mid-infrared excess \citep{2004MNRAS.353..991K}. Its magnetic field was detected by \cite{2008A&A...481L..99A}, who also noted its strong, variable He lines, suggesting that it may be a He-strong star. W\,601 is an X-ray source \citep{2014ApJS..215...10N}, and a radio synchrotron source \citep{2017MNRAS.465.2160K}, both likely indicative that the star hosts a detectable magnetosphere; the other possible explanation for enhanced X-rays and non-thermal radio emission -- colliding winds -- seems unlikely due to the relatively weak winds of B-type stars, and unneccessary due to the presence of a magnetic field.

The observations -- a large ESPaDOnS dataset -- are described in \S~\ref{sec:obs}. In \S~\ref{sec:binary} we show the evidence that W\,601 is a spectroscopic binary, and constrain the orbital properties of the system. Stellar parameters are revisted in the light of binarity in \S~\ref{sec:stellar_pars}. The magnetic analysis is presented in \S~\ref{sec:magnetometry}, where it is determined that the magnetic field belongs to the secondary. Spectroscopic variability is examined in \S~\ref{sec:halpha_em}, and in \S~\ref{sec:rot_em} the rotational period is determined from the magnetic and spectroscopic variations. The magnetic properties of the stars are derived in \S~\ref{sec:orm}. The magnetospheric properties of the system, and its possible future magnetic and rotational evolution, are discussed in \S~\ref{sec:discussion}, along with implications for the origins of fossil magnetic fields, their accompanying chemical abundance anomalies, and the status of the system as a Herbig star. Conclusions are summarized in \S~\ref{sec:conclusion}.

   \begin{figure*}
   \centering
   \includegraphics[width=0.95\textwidth]{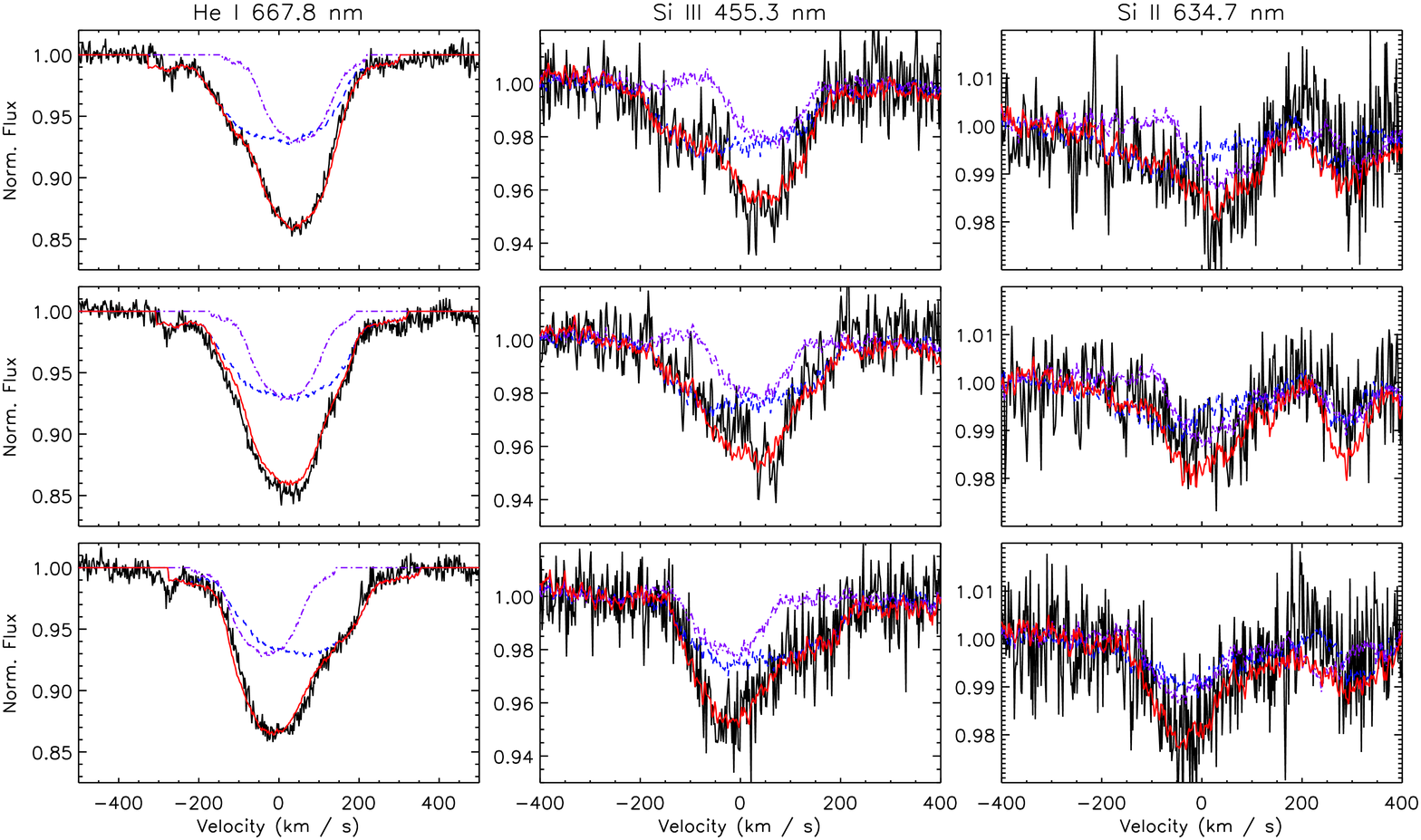} 
      \caption[]{Observed line profiles (black lines) and disentangled line profiles (dashed blue: primary; dot-dashed purple: secondary; solid red: cumlative) for 3 isolated spectral lines. Each row shows the same spectrum. The top, middle, and bottom rows respectively show spectra with the secondary at maximum positive radial velocity, the minimum difference in radial velocity between primary and secondary, and with the secondary at maximum negative radial velocity. Line profile variations in all lines can be reproduced assuming a two-star model with variable RVs.}
         \label{W601_HeI6678}
   \end{figure*}

   \begin{figure*}
   \centering
\begin{tabular}{ccc}
   \includegraphics[width=0.33\textwidth,trim=80 0 0 0]{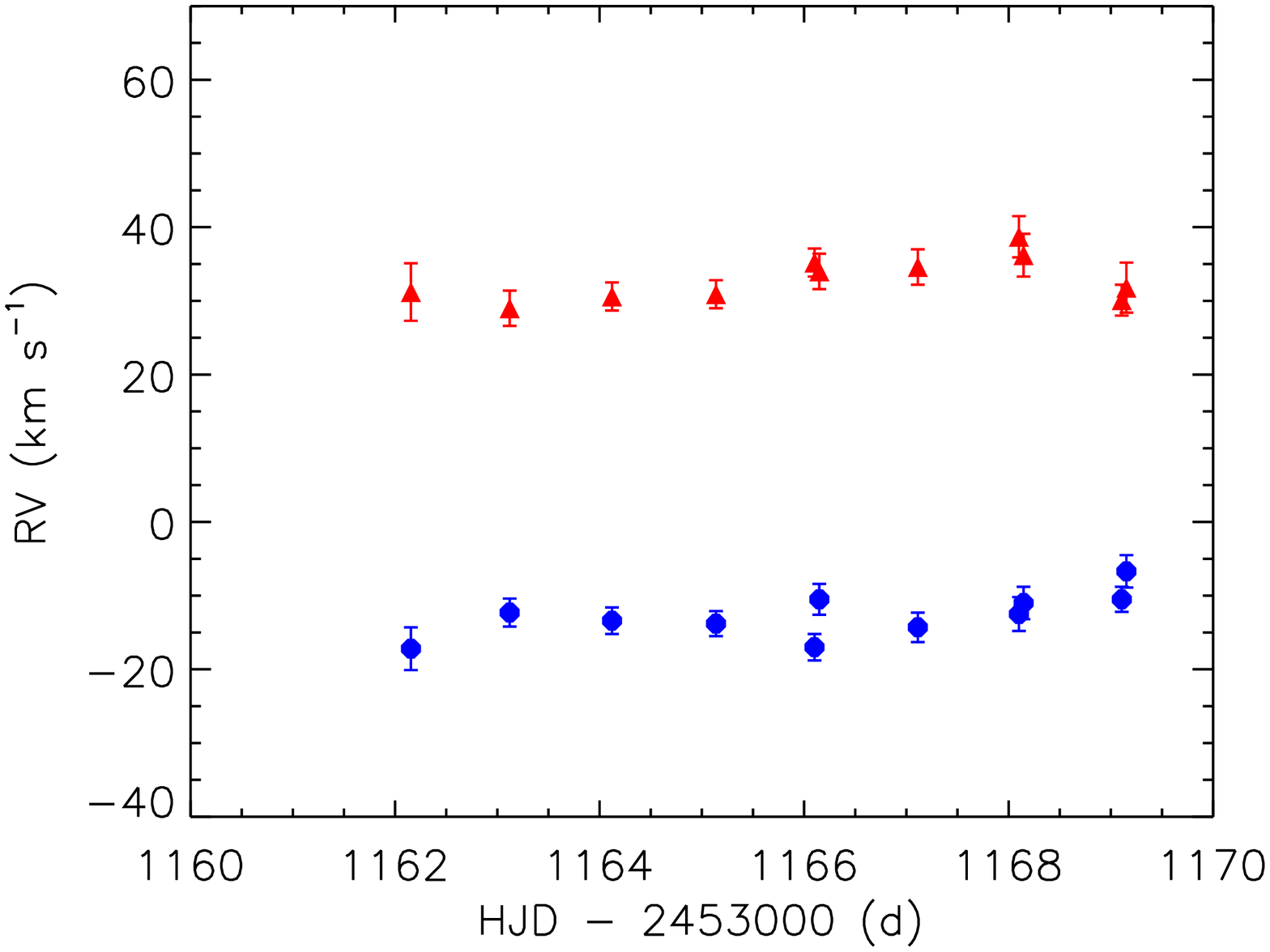} &
   \includegraphics[width=0.33\textwidth,trim=80 0 0 0]{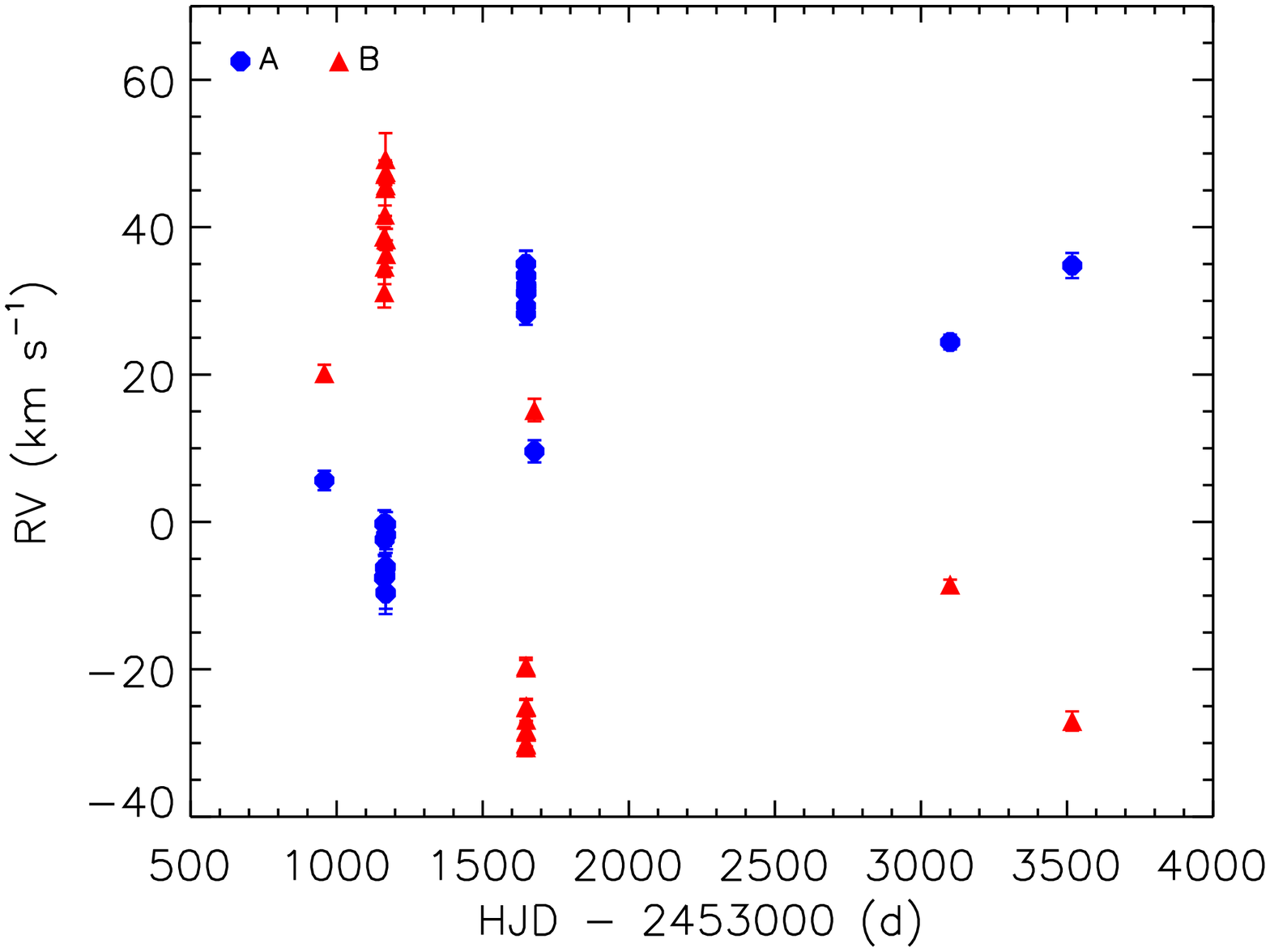} &
   \includegraphics[width=0.33\textwidth,trim=80 0 0 0]{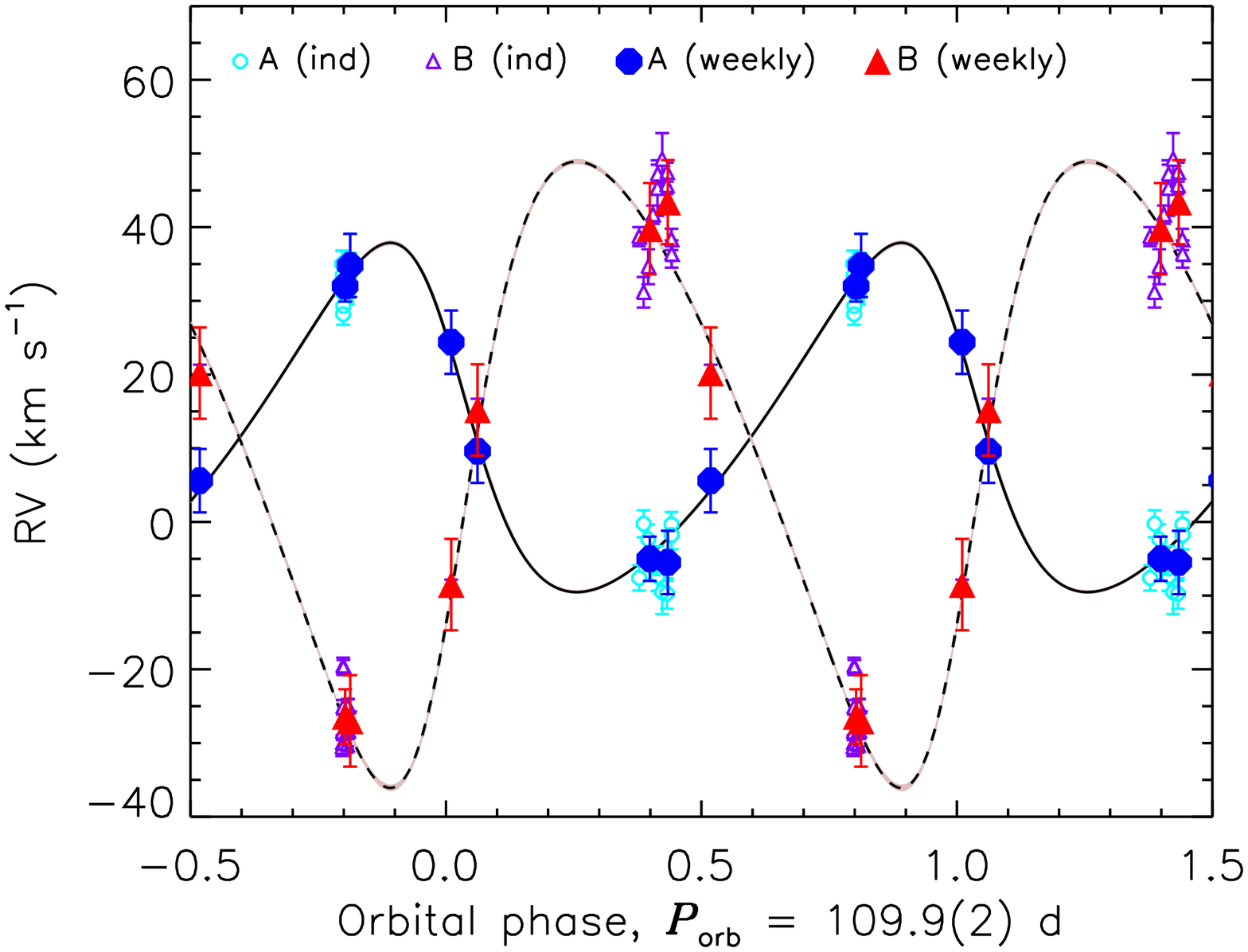} \\
\end{tabular}
      \caption[]{Radial velocities for the primary (blue circles) and secondary (red triangles) as a function of time (left, middle) and phased with the orbital period (right). The left panel zooms in on the epoch with the densest time sampling. In the right panel, curves and shaded regions indicate orbital models and uncertainties.}
         \label{W601_rv}
   \end{figure*}

   \begin{figure}
   \centering
   \includegraphics[width=0.45\textwidth]{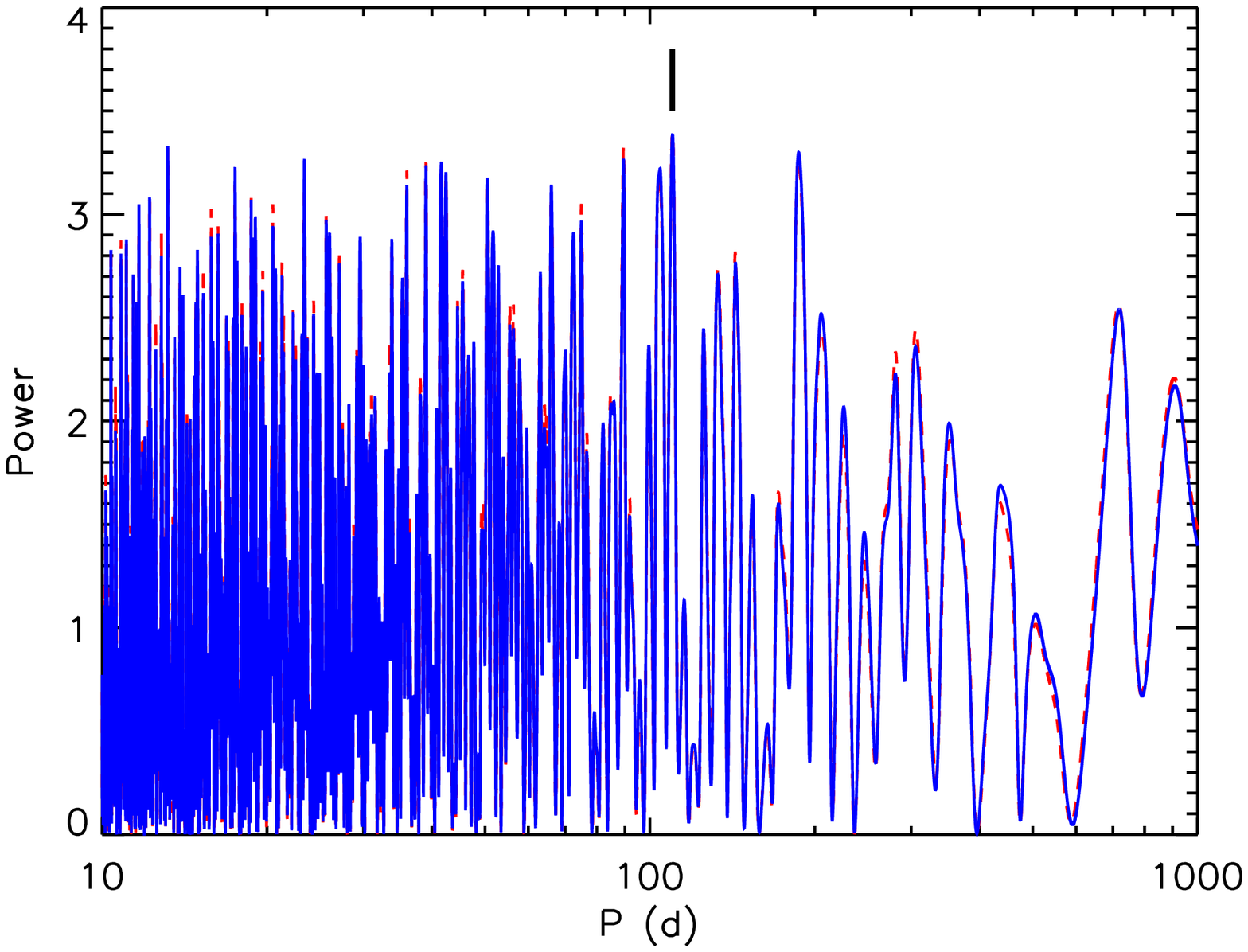} 
      \caption[]{Periodograms for the primary (solid blue) and secondary (dashed red) mean weekly RVs. The thick black line indicates the maximum power period. Note that there are several peaks of nearly equivalent significance; these are discussed further in the text.}
         \label{W601_periods}
   \end{figure}

\section{Observations}\label{sec:obs}

The dataset consists of 27 spectropolarimetric ESPaDOnS sequences obtained at the 3.6~m Canada-France-Hawaii Telescope (CFHT) from several observing programs (program codes 06BF15, 07AF06, 08AF16, 08AC13, 08BP14, 12AP13, and 13BC09), and in the context of the Magnetism in Massive Stars (MiMeS) Large Program. ESPaDOnS is a fibre-fed echelle spectropolarimeter with a spectral resolution $\lambda/\Delta\lambda \sim 65,000$ at 500 nm, and a wavelength coverage between about 370 nm and 1050 nm. Each observation consists of 4 polarized sub-exposures, which are combined to yield a circular polarization (Stokes $V$) spectrum and two diagnostic null $N$ spectra, with which anomalies in instrument behaviour can be detected. The reduction and analysis of ESPaDOnS data was described in detail by \cite{2016MNRAS.456....2W}. The data were reduced with the Libre-ESPRIT pipeline \citep{d1997}.

The observation log is given in Table \ref{obstab}. The mean peak signal-to-noise ($S/N$) per spectral pixel is about 200. 

Four of these observations were already reported by \cite{2008A&A...481L..99A} (one on 10/08/2006, two on 06/03/2007, and one on 09/03/07). The remaining observation are presented here for the first time.

\section{Binarity}\label{sec:binary}

Magnetic Bp stars exhibit several forms of line profile variation. The first, and most common in this class, are due to inhomogeneous surface chemical abundance patches, almost invariably exhibiting different patterns of variation for different elements \citep[e.g.][]{2015A&A...574A..79K,2015A&A...573A.123R,2016A&A...588A.138R,2017MNRAS.471..962S,2018A&A...609A..88R,2019A&A...621A..47K}. Many Bp stars also exhibit line profile variability due to pulsation or binarity. As is shown below, the line profile variations of W\,601 reported by \cite{2008A&A...481L..99A}, and interpted there as the signature of He spots, are in fact primarily due to binarity.

Close examination of the line profile variability shows that lines of different elements exhibit essentially the same pattern of variation (Fig.\ \ref{W601_HeI6678}), with absorption excesses appearing in the same part of the line regardless of element.  Since it is not expected that different elements will display the same abundance distributions, the similarity of the line profile variability between lines from different elements suggests that the source of the variation may instead be either binarity or pulsation. 

To test the binarity hypothesis, two-component models were fit to the strong He~{\sc i}~667.8~nm line using the paramaterized line profile fitting routine described by \cite{2017MNRAS.465.2432G}. This process yielded consistent fits for a broad-lined component (\vsini~$=173 \pm 14$~\kms) and a narrow-lined component (\vsini~$=105 \pm 3$~\kms), where the uncertainties are determined from the standard deviation of fits across the dataset. The radial velocities (RVs) of the two components obtained in the course of profile fitting anticorrelate with one another (Fig. \ref{W601_rv}, top). Since the broad-lined component has the smaller RV amplitude, we designate it as the primary, W\,601\,A, and the narrow-lined component as W\,601\,B. W\,601\,A has an RV semi-amplitude of about 20 \kms~and W\,601\,B has a semi-amplitude of about 30 \kms. There is no detectable RV variation on a time-scale of one to two days, however there is significant variation over a time-scale of weeks. This is in contrast to EW variations tracing rotational variability, which occur on a much shorter timescale (see \S~\ref{sec:halpha_em}).

The RVs and \vsini~values were then used to disentangle the line profiles of various spectral lines using the iterative method described by \cite{2006AA...448..283G}, as shown for the He~{\sc i}~667.8~nm, Si~{\sc iii}~455.3~nm, and Si~{\sc ii}~634.7~nm lines in Fig.\ \ref{W601_HeI6678}. As can be seen in Fig.\ \ref{W601_HeI6678}, the broad-lined component accounts for the majority of line absorption: 57\% of the absorption in He~{\sc i}~667.8~nm, 74\% in Si~{\sc iii}~455.3~nm, and 63\% in Si~{\sc ii}~634.7~nm. 

The anticorrelation of the RVs, and the fact that the larger RV amplitude is seen in the component with the smaller contribution to the line absorption, are all consistent with W\,601 being a double-lined spectroscopic binary (SB2). 

Period analyses of the RVs were performed using both the standard Fourier analysis package {\sc period04}\footnote{Available at \url{https://www.univie.ac.at/tops/Period04/}} \citep{2005CoAst.146...53L} and using the {\sc idl} routine {\sc periodogram.pro}\footnote{Available at \url{https://hesperia.gsfc.nasa.gov/ssw/gen/idl/util/periodogram.pro}} \citep{1986ApJ...302..757H}. Period uncertainties were determined analytically \citep{1976fats.book.....B}. Analysis of the full RV dataset yielded ambiguous results, with numerous peaks of similar amplitude between about 30 and 200 days. Since there is only minimal variation on timescales of a few days (Fig.\ \ref{W601_rv}, left panel), we analyzed mean weekly RVs in order to avoid biasing the periodogram results by the few highly sampled epochs. This strategy yielded a period of 109.9(2)~d from both the A and B RVs (see Fig.\ \ref{W601_periods}, where periodograms from {\sc periodogram.pro} are shown). {\sc period04} instead yielded a period of 104 d, however, the longer period obtained using {\sc periodogram.pro} provides a better phasing of the data (Fig.\ \ref{W601_rv}, right). The longer period furthermore yields a higher $S/N$ as evaluated using {\sc period04} itself (26 as compared to 5.5). We therefore adopt the period from {\sc periodogram.pro} as the most likely orbital period, although we note that there is enough ambiguity in the period that it should certainly be tested by further observation.

To obtain an orbital solution, we optimized synthetic RV curves against the mean weekly RVs using a Markov Chain Monte Carlo algorithm \citep[as described by ][]{2018MNRAS.475..839S,2019MNRAS.482.3950S}. The algorithm achieved rapid convergence, providing additional confidence in the period. The model fits to the RVs are shown in the right panel of Fig.\ \ref{W601_rv}, and the fit parameters are given in Table \ref{orbtab}. The orbit is mildly eccentric ($e = 0.24 \pm 0.01$), with a mass ratio $M_{\rm A}/M_{\rm B} = 1.79 \pm 0.04$. The $\chi^2$ of the fit shown in Fig.\ \ref{W601_rv} is 3.6. MCMC fitting of the 104 d period yielded a $\chi^2$ of 15.4, with a noticeably worse fit. The periodogram also has a peak at 186 d, nearly the same strength as the 109 d period; fitting with this period yielded a $\chi^2$ of 6.7, somewhat higher than the 109 d period. This indicates that the 109 d period is the most likely to be correct. Notably, while the 3 fits yield different eccentricities $e$ and velocity semi-amplitudes $K$, the systemic velocities $v_0$ and the mass ratios $M_{\rm A}/M_{\rm B} = K_{\rm B}/K_{\rm A}$ are essentially identical in all three cases, indicating that these parameters are likely robust against future revision of the orbital period.

\begin{table}
\centering
\caption[]{Orbital parameters: orbital period $P_{\rm orb}$; epoch of periastron $T_0$; eccentricity $e$; argument of periastron $\omega$; central velocity $v_0$; semi-amplitudes $K_{\rm A}$ and $K_{\rm B}$; mass ratio $M_{\rm A}/M_{\rm B}$; projected total mass $M\sin^3{i}$; projected component masses $M_{\rm A}\sin^3{i}$ and $M_{\rm B}\sin^3{i}$; and projected semi-major axis $a\sin{i}$.}
\label{orbtab}
\begin{tabular}{l r r}
\hline\hline
Parameter & Value & Uncertainty \\
\hline
$P_{\rm orb}~({\rm d})$ & 109.9 & 0.2 \\
$T_0~({\rm HJD})$ & 2453900.8 & 0.2 \\
$e$ & 0.234 & 0.008 \\
$\omega~(^\circ)$ &  64.6 & 0.9 \\
$v_0~({\rm km/s})$ & 11.4 & 0.1 \\
$K_{\rm A}~({\rm km/s})$ & 21.5 & 0.2 \\
$K_{\rm B}~({\rm km/s})$ & 41.8 & 0.4 \\
$M_{\rm A}/M_{\rm B}$ & 1.94 & 0.03 \\
$M\sin^3{i}~({\rm M_\odot})$ & 2.64 & 0.06 \\
$M_{\rm A}\sin^3{i}~({\rm M_\odot})$ & 1.75 & 0.04 \\
$M_{\rm B}\sin^3{i}~({\rm M_\odot})$ & 0.90 & 0.02 \\
$a\sin{i}~({\rm AU})$ & 0.621 & 0.005 \\
\hline\hline
\end{tabular}
\end{table}

\section{Stellar parameters}\label{sec:stellar_pars}

   \begin{figure*}
   \centering
   \includegraphics[width=0.95\textwidth, trim=100 50 50 0]{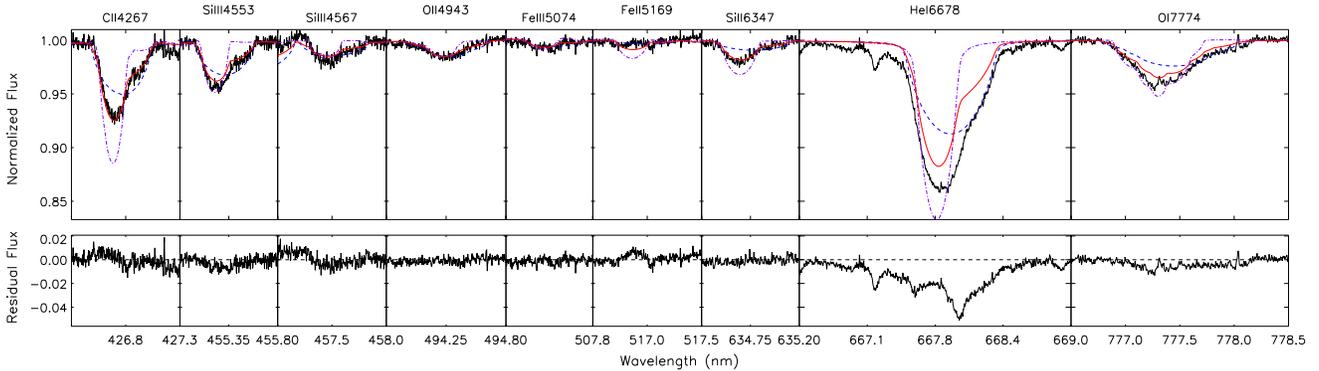}
      \caption[]{Comparison of the best-fit composite {\sc TLUSTY} spectrum (red lines) to the observed mean spectrum obtained via co-addition of observations with similar RVs (black lines). Contributions from the primary (dashed blue) and secondary (dot-dashed purple) show the intrinsic flux from each component, i.e.\ not scaled by the luminosity ratio. The bottom panel show the residuals after subtraction of the model. Note the poor fit to He~{\sc i}~667.8~nm, indicative of one of the stars being a He-strong star; this line was not used for parameter determination.}
         \label{tlusty_fit}
   \end{figure*}

   \begin{figure}
   \centering
   \includegraphics[width=0.45\textwidth]{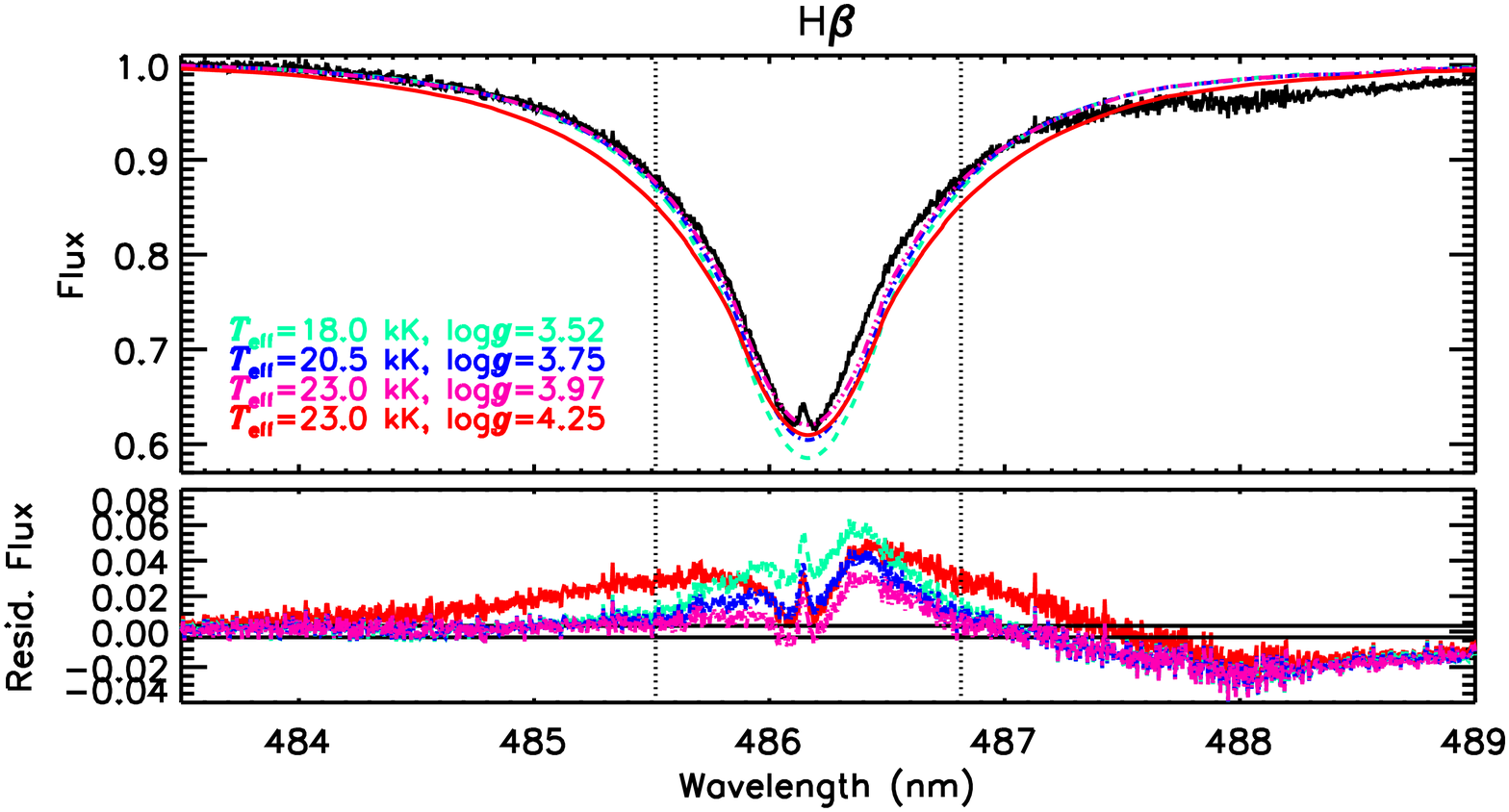}
   \includegraphics[width=0.45\textwidth]{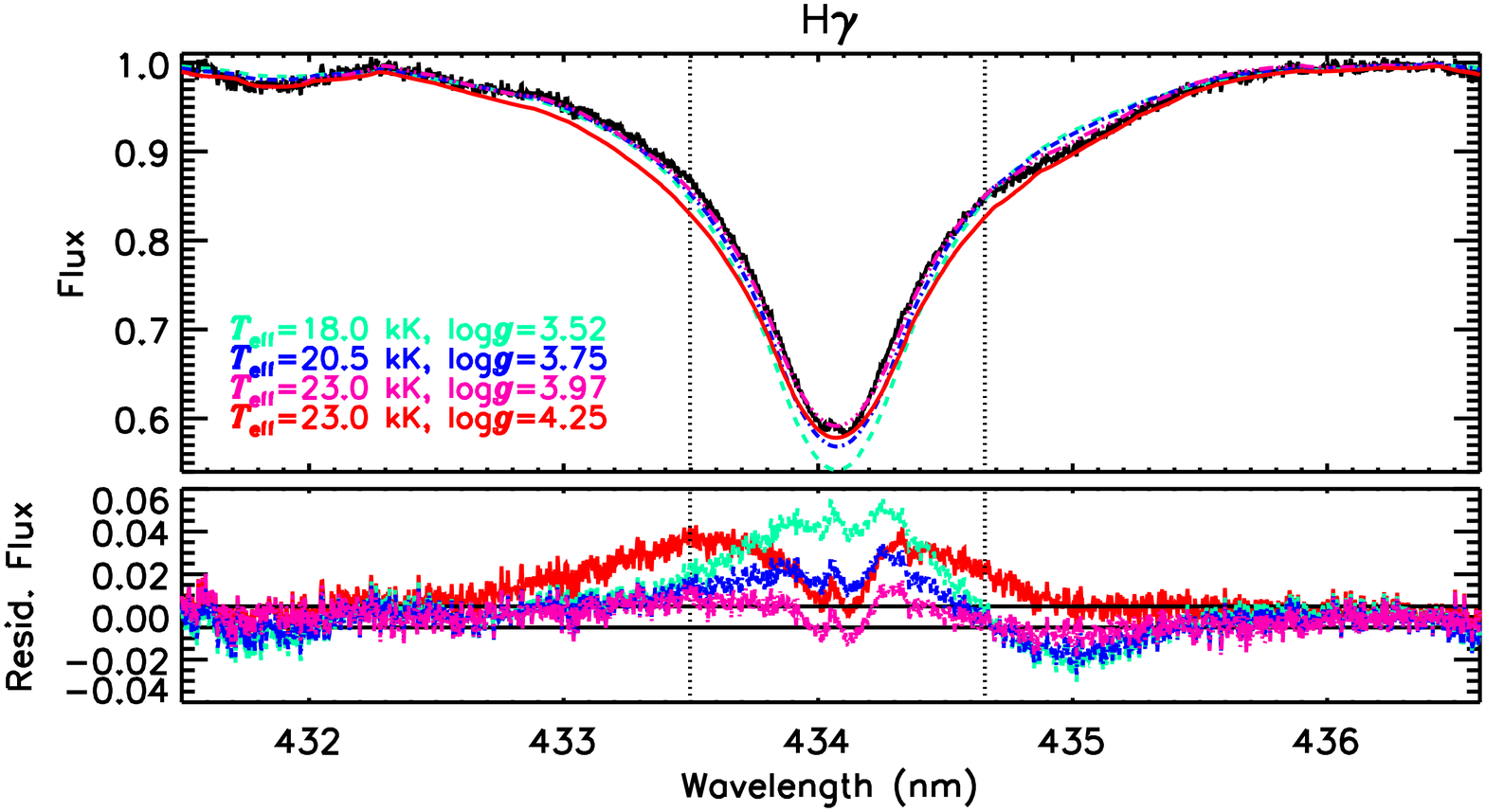} 
      \caption[]{Single component fits to the H$\beta$ and H$\gamma$ lines. Vertical dotted lines indicate regions excluded from the fit. Note the flat residuals outside the exclusion regions; residuals inside are due to circumstellar emission. Legends indicate the \teff~and $\log{g}$ of the corresponding models. Bottom sub-panels show the residual flux. Cyan, blue, and purple lines indicate best fit values for the given \teff; the red line shows $\log{g}=4.25$ for comparison. Note the much deeper wings of the $\log{g}=4.25$ model in comparison to the observed spectrum.}
         \label{logg}
   \end{figure}

   \begin{figure}
   \centering
   \includegraphics[width=0.45\textwidth,trim=50 0 75 100]{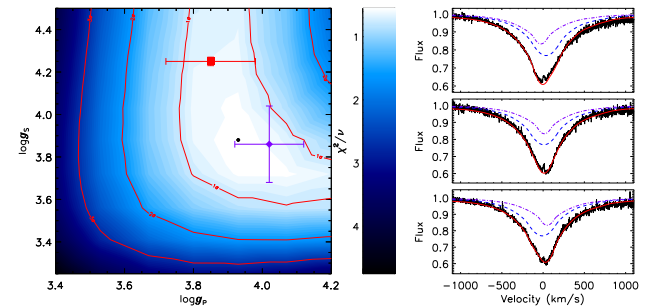} 
      \caption[]{Two-component fit to H$\beta$. The left panel shows the $\chi^2$ landscape (range indicated by colour bar), with the best-fitting model parameter indicated with black dots. Red contours indicate $\sigma$-levels. The diamond and square respectively indicate surface gravities inferred from main-sequence and pre-main-sequence evolutionary models (Fig.\ \protect\ref{physpar}). Right panels show the fits to H$\beta$, at (top to bottom), maximum primary RV, minimum RV separation, minimum primary RV (observed: black lines; dashed blue lines, primary; dot-dashed purple lines, secondary; solid red lines, combined flux).}
         \label{w601_bin_logg_fit}
   \end{figure}

   \begin{figure}
   \centering
   \includegraphics[width=0.45\textwidth]{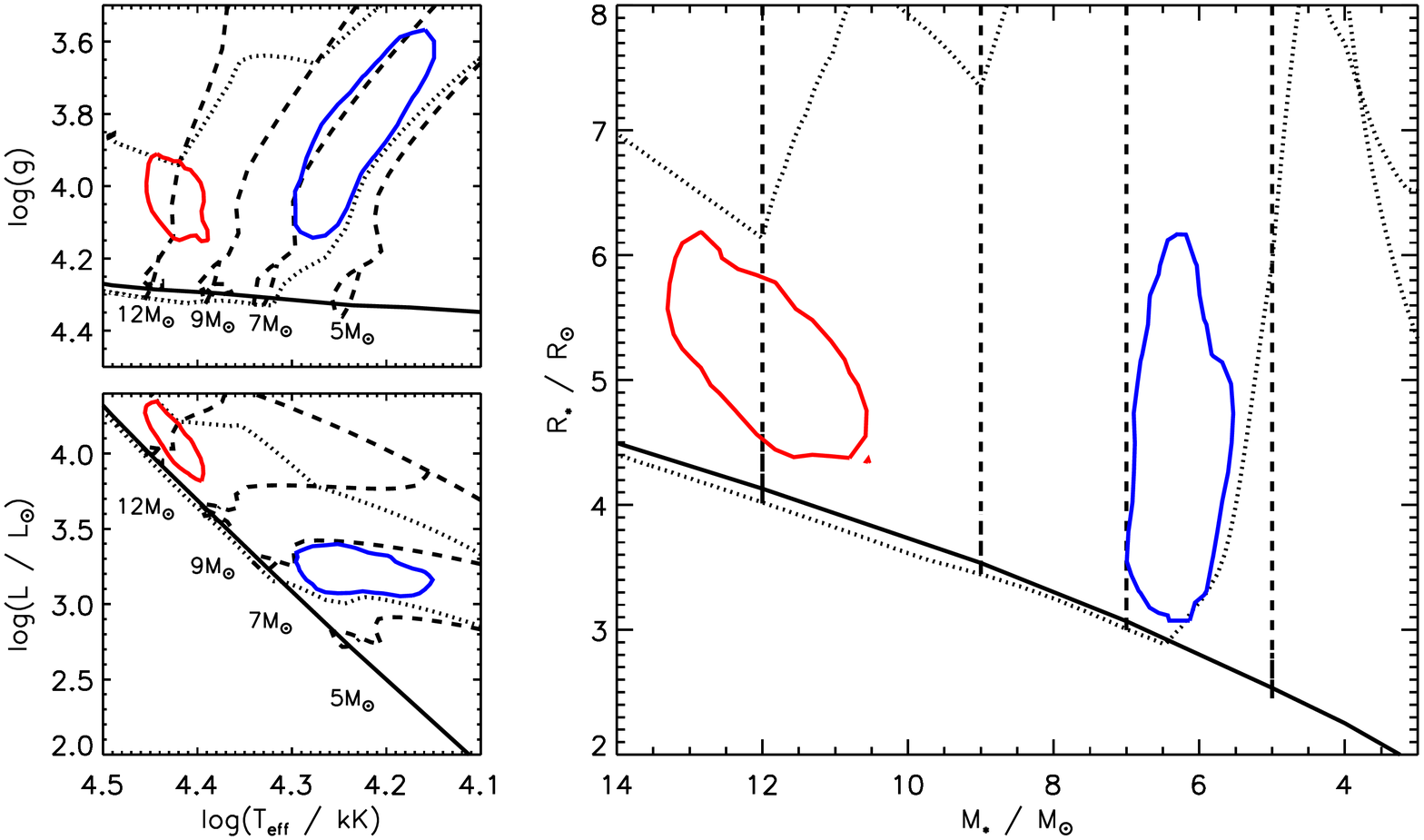} 
   \includegraphics[width=0.45\textwidth]{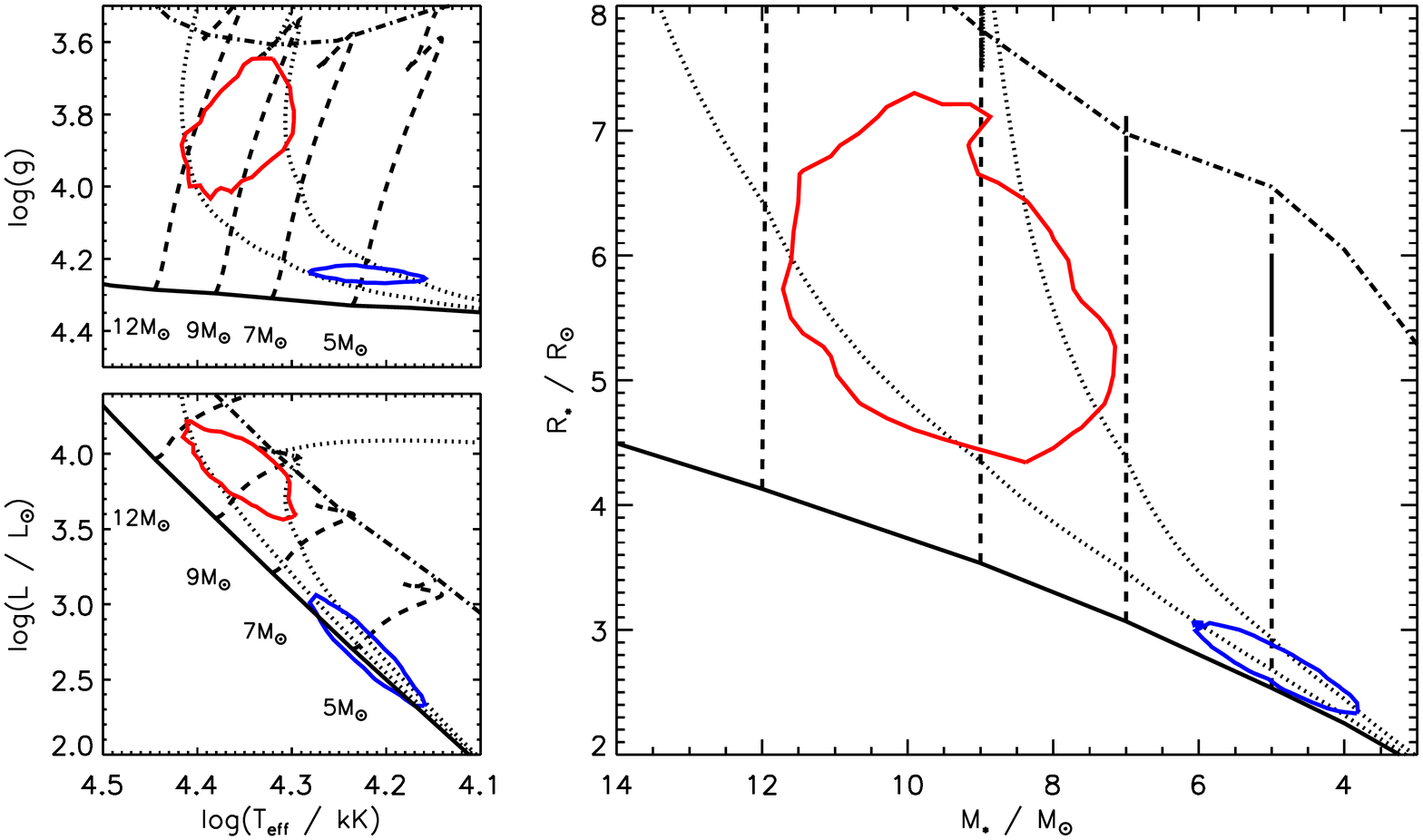} 
      \caption[]{Stellar parameters inferred from MC sampling of the HRD using pre-MS models \citep[][top panels]{Haemmerle2019} and MS models \citep[][bottom panels]{ekstrom2012}. Contours show 1$\sigma$ uncertainties (red: primary; blue: secondary). The solid line shows the Zero-Age Main Sequence; dashed and dotted lines respectively show evolutionary tracks and isochrones. For the PMS models, isochrones are shown for $\log{t} = 5.4$ and 5.7; for the MS models, isochrones for $\log{t} = 7.1$ and 7.5 are shown. A comparison between the $\log{g}$ values determined from evolutionary models to those measured via spectroscopic modelling is shown in Fig.\ \protect\ref{w601_bin_logg_fit}.}
         \label{physpar}
   \end{figure}

\begin{table}
\centering
\caption[]{Stellar parameters for the W\,601 components. $^a$: \protect\cite{2019ApJ...870...32K}; $^b$: using the anomalous extinction law determined by \protect\cite{2006A&A...457..265D}; $^c$: \protect\cite{nieva2013}. Pre-Main Sequence (PMS) models: \protect\cite{Haemmerle2019}; Main Sequence (MS) models: \protect\cite{ekstrom2012}.}
\label{partab}
\begin{tabular}{l c c}
\hline\hline
Parameter & A & B \\
\hline
$B~({\rm mag})$ & \multicolumn{2}{c}{11.11} \\
$V~({\rm mag})$ & \multicolumn{2}{c}{10.78} \\
$d~{\rm (kpc)}$$^a$ & \multicolumn{2}{c}{$1.74 \pm 0.13$} \\
$E(B - V)$ &  \multicolumn{2}{c}{$0.68 \pm 0.05$} \\ 
$A_{\rm V}~({\rm mag})$$^b$ & \multicolumn{2}{c}{$2.38 \pm 0.3$} \\
$M_V~({\rm mag})$ & $-3.0 \pm 0.4$ & $-1.7 \pm 0.3$ \\
${\rm BC}~({\rm mag})$$^c$ & $-2.6 \pm 0.1$ & $-1.6 \pm 0.2$ \\
$M_{\rm bol}~({\rm mag})$ & $-5.7 \pm 0.4$ & $-3.3 \pm 0.3$ \\
$T_{\rm eff}~({\rm kK})$ & $22 \pm 1$ & $19 \pm 1$ \\
$\log{g}~({\rm cm/s})$ & $3.9^{+0.3}_{-0.1}$ & $3.9^{+0.5}_{-0.3}$  \\
$\log{(L/L_\odot)}$ & $4.1 \pm 0.2$ & $3.2 \pm 0.1$ \\
$R_*~({\rm R_\odot})$ & $5.2 \pm 0.7$ & $4.3 \pm 0.9$ \\
$M_*~({\rm M_\odot})$ & $12 \pm 1$ & $6.2 \pm 0.4$ \\
$\log{(t/{\rm yr})}$ (PMS) & \multicolumn{2}{c}{$5.6 \pm 0.2$} \\
$\log{(t/{\rm yr})}$ (MS) & \multicolumn{2}{c}{$7.3 \pm 0.2$} \\
$i_{\rm orb}$ ($^\circ$) & \multicolumn{2}{c}{$31.5 \pm 0.9$} \\
$a$ (AU) & \multicolumn{2}{c}{$1.18 \pm 0.02$} \\
\hline\hline
\end{tabular}
\end{table}

Stellar parameters for W\,601 have previously been determined by \cite{1997A&AS..121..223D} ($\log{L/L_\odot} = 4.28$, \teff~$=23.5$~kK) and \cite{2006A&A...457..265D} ($\log{L/L_\odot} = 3.96 \pm 0.11$, \teff~$=22.4 \pm 2.7$ kK, and $\log{g} = 3.85$), where the different luminosities arise from different assumptions regarding distance and extinction. Since these parameters were determined under the assumption of a single star, they need to be revisited. 

\subsection{Effective Temperature}

As a first pass to estimate the \teff~of the two components, we used EW ratios of Si~{\sc ii}~634.7~nm and Si~{\sc iii}~455.3~nm obtained from disentangled line profiles (Fig.\ \ref{W601_HeI6678}). While surface abundances are by definition affected by the chemical spots expected for a magnetic chemically peculiar star, line strength ratios for different ions of the same element are not affected by this \citep[since it is the abundance, but not the \teff, that varies across the stellar surface; e.g.][]{2015MNRAS.449.3945S,2019MNRAS.485.1508S}. Comparison of the weighted mean EW ratios calculated from all spectra to theoretical values obtained from the NLTE {\sc TLUSTY} BSTAR2006 library of synthetic spectra \citep{lanzhubeny2007}, using $3.5 < \log{g} < 4.25$, yields \teff$_{\rm A} = 22.5 \pm 1.5$ kK and \teff$_{\rm B} = 20.5 \pm 1.5$ kK. While the \teff~of the two components overlaps within uncertainty, B is definitely cooler than A, since the contribution of A is higher in Si~{\sc iii} as compared to Si~{\sc ii}. 

A second analysis is demonstrated in Fig.\ \ref{tlusty_fit}, where we used a grid of synthetic spectra calculated from {\sc TLUSTY} models, with $15 < T_{\rm eff} < 25$~kK and $3.0 < \log{g} < 4.5$ for each star. The {\sc tlusty} spectra were calculated with 2 \kms~of microturbulence and solar metallicity. The radius ratio $R_{\rm A}/R_{\rm B}$ was allowed to vary as a free parameter between 0.5 and 5, where $R_{\rm A}/R_{\rm B}$ is a factor used to scale the contributions of the A and B components to the composite spectrum. $v\sin{i}_{\rm A} = 173$~\kms, $v\sin{i}_{\rm B} = 94$~\kms, ${\rm RV}_{\rm A} = 35$~\kms, and ${\rm RV}_{\rm B} = -23$~\kms~were fixed. The grid was compared to a mean spectrum created by combining the ESPaDOnS observations obtained in 03/2007, since the RVs of the two components are basically constant during the 1 week of observations (see the left panel of Fig.\ \ref{W601_rv}. The standard deviation in RV is 3.5 \kms~and 6~\kms~for W\,601\,A and B, respectively, in both cases comparable to the uncertainty in \vsini; co-addition of spectra therefore should not lead to significant additional line broadening)., and this is the epoch with the largest number of observations. The peak $S/N$ per 1.8 \kms~spectral pixel of the mean 2007 spectrum is 670. The analysis was performed on a selection of strong lines, shown in Fig.\ \ref{tlusty_fit}, with a focus on chemical species for which two ionization levels are present in the spectrum (O~{\sc i} and {\sc ii}; Si~{\sc ii} and {\sc iii}; and Fe~{\sc ii} and {\sc iii}). The results of this analysis are $T_{\rm eff,A} = 22 \pm 1$~kK, $T_{\rm eff,B} = 19 \pm 1$~kK, $\log{g} = 3.75 \pm 0.25$ for both stars, and $R_{\rm A}/R_{\rm B} = 1.1 \pm 0.1$. 

The results of the spectroscopic fit are consistent with those of the EW ratio analysis, but use a larger number of lines, and yield more precise results; we therefore adopt the spectroscopic fit results for the \teff. The surface gravity is revisited below using Balmer lines, as these are more sensitive to $\log{g}$.

\cite{2008A&A...481L..99A} classified W\,601 as a He-strong star because its He lines were much stronger than solar abundance models predict, as is usually the case for strongly magnetic stars in this \teff~regime. As can be seen in Fig.\ \ref{tlusty_fit}, a binary spectroscopic model with solar abundances also results in He lines much weaker than observed. For this reason, He lines were not included in the \teff~determination.

\subsection{Surface Gravity}

To refine the determination of $\log{g}$ we examined the H$\beta$ and H$\gamma$ lines. Since there is emission in H$\alpha$, there will also be (weaker) emission in the higher-numbered H Balmer lines. In order to mitigate the influence of this emission on the Balmer wings, a mean spectrum was created from the observations exhibiting the smallest amount of H$\alpha$ emission. The mean spectra were obtained from merged and normalized spectral orders in the same fashion as described by \cite{2019MNRAS.485.1508S}. Fits were performed outside of $\pm 200$~\kms~i.e.\ excluding the rotationally broadened core of the primary. While the H$\alpha$ emission can extend out to $\pm$700~\kms, in the case of the low-emission mean spectrum the significant emission is contained within the $\pm$200~\kms~exclusion range, as can be verified from the residuals in Fig.\ \ref{logg}. Rotationally broadened synthetic spectra from the BSTAR2006 library were utilized for the fits, with $\chi^2$ minima being determined for 3 values of \teff~spanning the range of \teff~of the primary and secondary. Here we have made the assumption that a single-star model can yield reasonable results, given both the low amplitude of the RV variation and the similar \teff~of the two stars. As can be seen from Fig.\ \ref{logg}, H$\beta$ and H$\gamma$ return similar values for $\log{g}$, $3.68 \pm 0.22$ and $3.75 \pm 0.23$ (cgs) respectively. 

We next performed a two-component fit to H$\beta$ accounting for the contributions and RV variation of both stars. Here the same method was adopted as by \cite{2019MNRAS.485.1508S}, with the radius of the secondary constrained by the orbital mass ratio and $\log{g}$. As before, only observations exhibiting minimal emission, confined to the central $\pm 200$~\kms~region of the line, were utilized, and this velocity range was excluded from the fit. The results are shown in Fig.\ \ref{w601_bin_logg_fit}, and are consistent with the single-component fit: $\log{g_{\rm A}} = 3.9^{+0.3}_{-0.1}$, and $\log{g_{\rm B}} = 3.9^{+0.5}_{-0.3}$.

\subsection{Fundamental Parameters and Evolutionary Status}

W\,601 has been reported as a pre-Main Sequence (PMS) star \citep{2008A&A...489..459M,2008A&A...481L..99A}. As its status as a classical Herbig Ae/Be star is doubtful due to the nature of its H$\alpha$ emission (see \S~\ref{sec:halpha_em} and \S~\ref{sec:rot_em}), it is not clear at this stage in the analysis whether the star is on the Main Sequence (MS) or the pre-MS, since its H$\alpha$ emission does not appear to be from an accretion disk. Here we seek to determine whether the star is on the PMS or on the MS. The low measured surface gravity is potentially consistent with two scenarios: either the stars are on the PMS, or the primary has evolved towards the terminal age MS (TAMS). 

As a first step, it is necessary to know whether or not W\,601 is a member of the young NGC\,6611 cluster, since if it is a member this establishes an upper limit on the star's age, and additionally gives better constraints on the distance than are available from the Gaia parallax of the individual star. \cite{2018A&A...618A..93C} identified W\,601 as a member of NGC\,6611 with an 80\% probability using Gaia DR2 parallaxes and proper motions \citep{2018A&A...616A...1G}. The star's DR2 parallax and proper motions are furthemore below the 3 and 5$\sigma$ rejection thresholds used by \cite{2019ApJ...870...32K} to determine cluster membership (respectively, differing by about 2.3$\sigma$ and 2.8$\sigma$ from the mean cluster values, using the smaller cluster uncertainty). The Gaia early Data Release 3 parallax and proper motions of W\,601 are furthermore identical within uncertainty to the mean cluster values determined by \cite{2019ApJ...870...32K} on the basis of Gaia DR2 values. W\,601's systemic velocity, $v_0 = 11.4 \pm 0.2$~\kms, is also consistent with the mean cluster radial velocity of $10 \pm 8$~\kms~determined from the VLT-FLAMES survey by \cite{2005A&A...437..467E}. We conclude that W\,601 is very likely to be a member of NGC\,6611.

The mean Gaia parallax of the cluster is $\pi_{\rm cl} = 0.57 \pm 0.04$~mas, corresponding to a distance of $d = 1740^{+130}_{-120}$~pc \citep{2019ApJ...870...32K}. From isochrone fitting of the MS turnoff, the NGC\,6611 cluster is between 0 and 6 Myr in age according to \cite{1997A&AS..121..223D}, while \cite{2006A&A...457..265D} found an age of $3 \pm 1$~Myr using VLT-FLAMES data. \cite{1993AJ....106.1906H} found ages of 0.25 Myr to 1 Myr for young stellar objects (YSOs) below 8 \msun, and determined an average age of $2 \pm 1$~Myr for stars above 9 \msun. \cite{2018MNRAS.476.1213G} found the cluster's YSOs to be between 1 and 2.6 Myr. All studies seem to agree that the oldest stars in the cluster are no more than 6 Myr in age, and that the cluster also contains ongoing star formation. 

Measurements of extinction in NGC\,6611 are widely variable across the region, ranging from $A_V =1.7$ to 7.2 \citep[e.g.][]{2006A&A...457..265D,2007A&A...462..245G,2018A&A...613A...9M}. There is furthermore evidence of a non-standard reddening law of between $R_V=3.5$ and 3.9 \citep[e.g.][]{1993AJ....106.1906H,2004MNRAS.353..991K,2007A&A...462..245G,2018A&A...613A...9M}. To constrain the extinction and reddening of W\,601, we utilized the star's $UBVJHK$ photometry \citep{2000A&A...355L..27H,2007AJ....133.1092W,cutri2003}, and de-reddened these using the {\sc idl} program {\sc fm\_unred.pro} \citep{1999ApJ...525.1011F}, with reddening $0 < E(B-V) < 1$ and reddening law $3 < R_V < 5$. The dereddened colours $U-B$, $B-V$, $V-J$, $V-H$ and $V-K$ were then compared to the intrinsic colours from the empirical PMS calibration determined by \cite{2013ApJS..208....9P}. The best match was obtained for $E(B-V) = 0.55$ and $R_V = 4.37$, yielding $A_V = 2.39$. The reddening law is steeper than obtained for other stars in the cluster, however fixing $R_V=3.5$ yields $E(B-V) = 0.68$, with $A_V$ almost unchanged. These results are consistent with those obtained by previous examinations of the star \citep[e.g.][]{2004MNRAS.353..991K,2006A&A...457..265D}.

In order to constrain the luminosity and the other fundamental stellar parameters, the binary Monte Carlo (MC) Hertzsprung-Russell Diagram (HRD) sampler described by \cite{2019MNRAS.488...64P} was utilized. This algorithm takes as priors the photometric properties of the system ($V$ magnitude, extinction, parallax), the orbital mass ratio (Table \ref{orbtab}), and the individual effective temperatures and surface gravities, and infers the luminosities, masses, ages, and radii via interpolation through evolutionary models under the assumption that the two components are coeval (but not interacting) \citep{1994MNRAS.271..999B}. Bolometric corrections for both stars were obtained from the calibration of \cite{nieva2013}. The \teff~$-\log{g}$ diagram is sampled with points drawn from gaussian distributions matching \teff~and $\log{g}$ values of the two stars, and are probabilistically rejected if the inferred mass ratio, absolute $V$ mag, and age differ from target values drawn from gaussian distributions with standard deviations set respectively by the measured uncertainty in the mass ratio, the combined uncertainty in distance modulus and extinction, and an arbitrary tolerance in $\log{t}$. The results are shown in Fig.\ \ref{physpar}, and the parameters are given in Table \ref{partab}.

   \begin{figure*}
   \centering
   \includegraphics[width=18.5cm, trim = 50 20 0 0]{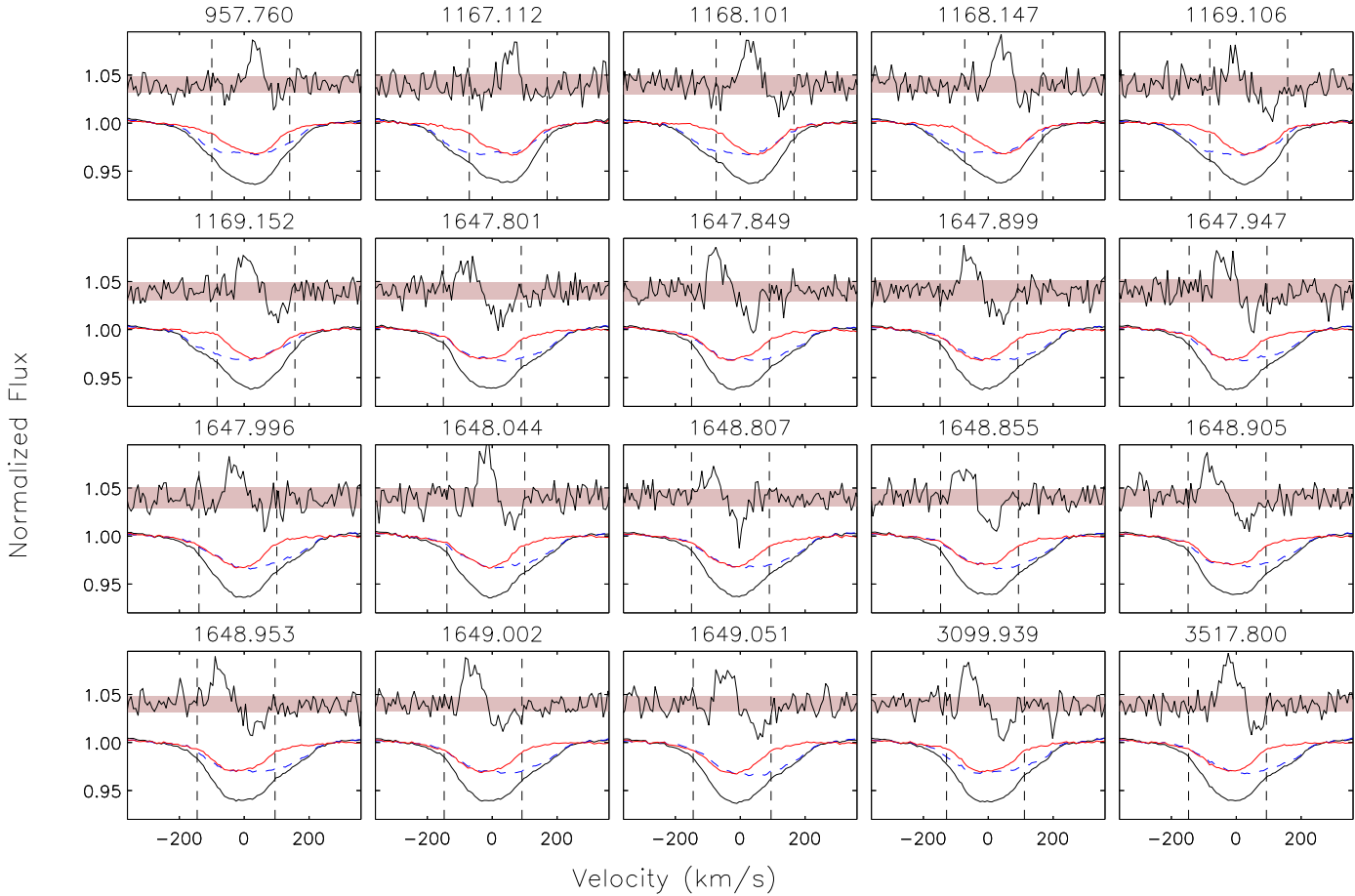} 
      \caption[]{LSD profiles. Each panel is labelled with HJD$-2453000$. Only LSD profiles yielding at least a marginal detection in Stokes $V$ are shown. In each panel, the full Stokes $I$ profile is shown with black lines. The disentangled Stokes $I$ profiles of W\,601\,A and B are respectively shown with dashed blue and solid red lines. Above Stokes $I$ is the Stokes $V$ profile. The shaded region indicates the mean uncertainty in Stokes $V$. Vertical dashed lines indicate the integration limits used for measuring the FAP and \bz. Note that Stokes $V$ is confined within the line profile of the secondary, and tracks its RV variation. It is therefore the secondary that is the magnetic star.}
         \label{W601_lsd}
   \end{figure*}

We determined stellar parameters using two sets of evolutionary models, the rotating MS models published by \cite{ekstrom2012} (i.e., those with an initial rotation velocity of 0.4 of the critical value; bottom panels of Fig.\ \ref{physpar}), and the rotating PMS models developed by \cite{Haemmerle2019} (top panels of Fig.\ \ref{physpar}), both of them calculated using the Geneva 1D stellar evolution code. The MS models start at the ZAMS, whereas the PMS models start at the birthline. For the PMS models an age tolerance of $\log{t}=0.15$ was adopted for the MC sampling, in order to reflect the possibility that very young stars might not be perfectly coeval; for the MS models a tolerance of 0.05 was used, ensuring any accepted points would lie on the same isochrone. 

PMS models yield an age of $\log{(t/{\rm yr})} = 5.6 \pm 0.1$, while the MS models yield an age of $\log{t} = 7.3 \pm 0.2$. Only the age derived from PMS evolutionary models is consistent with the 6 Myr ($\log{t} = 6.8$) upper limit on the cluster age. 

The MS models predict that W\,601\,A has a larger radius than W\,601\,B, whereas the PMS models yield similar radii (compare the top and bottom right panels of Fig.\ \ref{physpar}). The ratio of radii from PMS MC parameter determination ($R_{\rm A}/R_{\rm B} = 1.2 \pm 0.3$) is consistent with results from spectral modelling ($R_{\rm A}/R_{\rm B} = 1.1 \pm 0.1$). Using MS models, the radii ratio is $R_{\rm A}/R_{\rm B} = 2.1 \pm 0.3$, which is not consistent with the spectroscopy. 

While the MC sampler initially draws from distributions in $\log{g}$ and \teff~as determined from spectroscopic measurements, due to the various rejection criteria the posterior distributions of $\log{g}$ do not necessarily resemble the input distributions. MS models can only maintain coevality if $\log{g}_{\rm A} = 3.85 \pm 0.12$ and $\log{g}_{\rm S} = 4.24 \pm 0.02$, i.e.\ if the primary has evolved about halfway towards the TAMS, the secondary should still be very close to the zero-age MS (ZAMS). Conversely, PMS models yield $\log{g}_{\rm A} = 4.06 \pm 0.10$ and $\log{g}_{\rm B} = 3.86 \pm 0.18$. A comparison to the spectroscopic measurement of $\log{g}$ is provided on the $\chi^2$ map in Fig.\ \ref{w601_bin_logg_fit}. Both the MS and PMS results are consistent with the best-fit value within the 1$\sigma$ contours, although the PMS results are closest to the $\chi^2$ minima.

PMS models yield a better match to the measured surface gravities, radius ratio, and the age of the NGC\,6611 cluster. We therefore conclude that the scenario with the greatest consistency is that both components of W\,601 are still contracting towards the ZAMS. It is additionally notable that the H$\alpha$ line displays nebular emission (\S~\ref{sec:halpha_em}), suggesting the system is still partially embedded in the nebula.

As can be seen in the top panels of Fig.\ \ref{physpar}, PMS models indicate that the ages of W\,601\,A and B differ by about $\log{t} = 0.2$, or about 200 kyr at the inferred age of the system. Reducing the age tolerance does not change this result; in fact an age tolerance of $\log{t}=0.15$ is the minimum necessary to keep the rejection rate reasonably small. This may indicate that the two components are not precisely coeval, but that the primary formed before the secondary.

From the stellar masses, the MC parameter analysis additionally yields an orbital inclination $i_{\rm orb} = 31.5 \pm 0.9^\circ$ and thus a semi-major axis $a = 1.18 \pm 0.02$~AU. 

\section{Magnetometry}\label{sec:magnetometry}

In order to increase the $S/N$ of the line profile from which the magnetic field is measured, Least Squares Deconvolution \citep[LSD;][]{d1997,koch2010} profiles were extracted from the ESPaDOnS spectra. We used a line mask obtained with an `extract stellar' request from the Vienna Atomic Line Database \citep[VALD3;][]{piskunov1995, ryabchikova1997, kupka1999, kupka2000,2015PhyS...90e4005R} for a 19 kK solar metallicity star, with $\log{g}=3.8$, and a line depth threshold of 0.1. The mask was cleaned in the usual fashion in order to remove contamination from H Balmer, telluric, and interstellar lines \cite[e.g.][]{2018MNRAS.475.5144S}. In addition to the intersellar Ca lines, W\,601's spectrum contains several Diffuse Interstellar Bands (DIBs) which were also removed from the line mask. DIBs were identified by eye, including but not necessarily limited to lines with approximate wavelengths of 472 nm, 476 nm, 496 nm, 523 nm, 540 nm, 541 nm, 549 nm, 551 nm, 554 nm, 578 nm, 579 nm, 585 nm, 661 nm, and 666 nm. 

Due to the large \vsini~and low $S/N$ of the data, He lines were left in the mask. The resulting LSD profiles were scaled to a line depth of 0.1, a mean wavelength of 500 nm, and a mean Land\'e factor of 1.2. To further increase the per-pixel $S/N$ a velocity pixel width of 7.2 \kms~was used. A Tikhonov regularization factor of 0.2 was employed in order to suppress noise arising from the deconvolution process \citep{koch2010}. Following extraction, the LSD profiles were iteratively disentangled using the same procedure as adopted in \S~\ref{sec:binary} \citep[e.g.][]{2006AA...448..283G}. The results are shown in Fig.\ \ref{W601_lsd}. The Zeeman signature is confined within the line profile of W\,601\,B, and tracks the radial velocity variation of this component; thus, it is W\,601\,B that hosts the magnetic field. Further magnetic analysis was therefore conducted using the disentangled Stokes $I$ profiles of W\,601\,B. Since the non-magnetic star only contributes noise to Stokes $V$, flux dilution does not affect magnetic measurements so long as the disentangled spectrum of the magnetic component is used \citep[see e.g.][]{2019MNRAS.489.5669P}.

Observations were classified as definite, marginal, or non-detections (DD, MD, or ND) according to the False Alarm Probabilities (FAPs) measured inside the line profile, using the method and criteria described by \cite{1992AA...265..669D,d1997}. The integration ranges used for measuring the FAP are shown in Fig.\ \ref{W601_lsd}, and detection flags for individual stars are given in Table \ref{obstab}. Of the 27 total observations, 18 yield DDs, 2 MDs, and 7 NDs. FAPs measured from the null $N$ spectra are uniformly non-detections. 

We measured the strength of the magnetic field via the disk-averaged longitudinal magnetic field \citep[e.g.][]{mat1989,wade2000}, using the same integration ranges as used for measuring the FAP (i.e.\ about $\pm$120~\kms~centred on the RV of W\,601\,B). These measurements, as well as the \nz~measurements obtained from the $N$ profiles, are given in Table \ref{obstab}. 

While the $N$ profiles are all NDs, the \nz~measurements show a systemetic bias towards negative values. Close examination revealed that several of the Stokes $V$ and $N$ continua of several of the observations are offset from the expected value of 0. This may be a consequence of the ratio of continuum flux in the two polarization beams changing as the retarder rotates \citep{2012A&A...538A.129B}, likely exacerbated by the low $S/N$ of the observations. Re-adjusting the continua of the LSD Stokes $V$ and $N$ profiles to null corrected this issue, yielding \nz~scattered evenly about 0. The amplitude of \bz~was slightly increased, although not outside of the error bars.

\bz~ranges from about $-1$ to $+2$ kG, with a mean uncertainty of $360$ G. Even without knowing the rotation period or magnetic configuration, the maximum value of \bz~indicates that the surface magnetic dipole strength $B_{\rm d}$ must be at least 7 kG (under the simplest assumption of a dipolar surface magnetic field, in which case the lower limit on $B_{\rm d}$ is about 3.5$\times$ the maximum value of \bz). Since both positive and negative values of \bz~are measured, both magnetic poles must come into view over the course of a rotational cycle.

\section{Line Profile Variability}\label{sec:halpha_em}

   \begin{figure}
   \centering
   \includegraphics[width=0.45\textwidth]{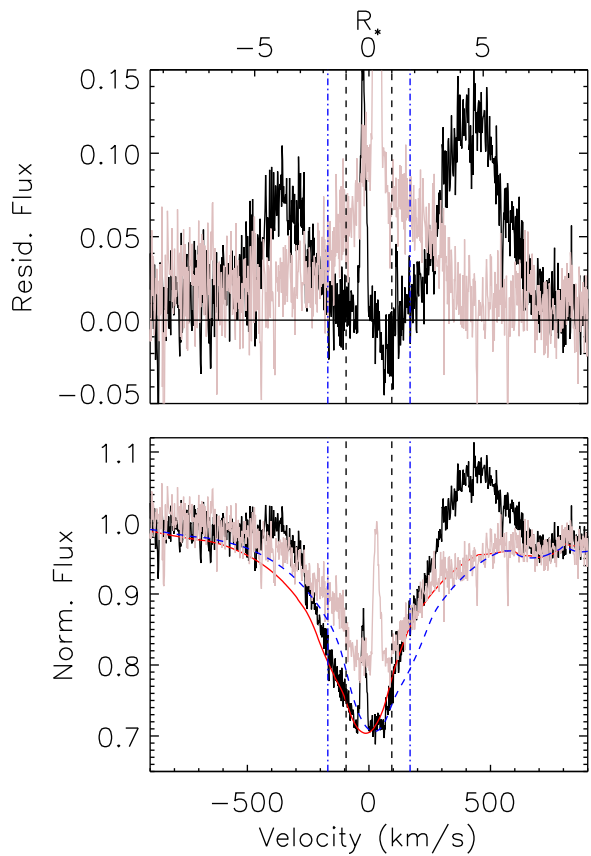} 
      \caption[]{{\em Bottom}: H$\alpha$ profiles at minimum emission (grey) and maximum emission (black), with respective synthetic binary profiles (dashed blue, solid red). Dashed black vertical lines indicate $\pm$\vsini~for W\,601\,B; dot-dashed blue vertical lines indicate $\pm$\rk. The profiles have been shifted to the rest-frame of the secondary. The thin emission feature in the centre of the line is nebular emission. {\em Top}: Residual flux after subtracting synthetic from observed spectra. The double-humped emission, peaking at several times \vsini, is a characteristic signature of a centrifugal magnetosphere. Note that the top axis is in units of stellar radii, reflecting the assumption that the emission is produced by magnetically confined material in corotation with the star.}
         \label{W601_halpha_minmax}
   \end{figure}

W\,601 was originally classified as a Herbig Be star on the basis of its youth and H$\alpha$ emission \citep{2008A&A...489..459M,2008A&A...481L..99A}. T Tauri stars and Herbig Ae/Be stars generally show variable emission in multiple lines, not only H$\alpha$ \citep{2011A&A...529A..34M}. The morphologies and variability patterns of the line emission of T Tauri stars are explained by the magnetospheric accretion model \citep{2007A&A...463.1017B}, in which the inner disk material is locked to and therefore in corotation with the star's magnetic field, leading to the formation of a gap between the photosphere and the inner rim of the disk. Linear H$\alpha$ spectropolarimetry confirms the existence of this gap in both T Tauri stars and Herbig Ae stars \citep{2005MNRAS.359.1049V}, and the line emission of at least one magnetic Herbig Ae star is consistent with magnetospheric accretion \citep{2016A&A...592A..50S}. However, the line emission of early-type Herbig Be stars is generally not consistent with magnetospheric accretion \citep{2011A&A...535A..99M,2020MNRAS.493..234W}, which is not surprising given the very low incidence of detected magnetic fields in this population.

The presence of a strong magnetic field and evidently rapid rotation may indicate that W\,601 may be the first Herbig Be star with H$\alpha$ emission consistent with magnetospheric accretion. However, the absence of the expected emission in other lines \citep{2011A&A...529A..34M} suggests that the star's H$\alpha$ profile may instead be consistent with an origin in a wind-fed `Centrifugal Magnetosphere' \citep[CM;][]{lb1978,petit2013}, since rapidly rotating, strongly magnetic early B-type stars almost invariably possess CMs detectable in H$\alpha$ \citep{2019MNRAS.490..274S}. 

Inspection of the H$\alpha$ line indicates that the morphology of the H$\alpha$ emission is consistent with a CM, i.e.\ a double-humped emission profile with the peak emission occuring at velocities equal to several times the projected rotational velocity \vsini, as can be seen in Fig.\ \ref{W601_halpha_minmax}. This shape arises due to rigid corotation of magnetically confined plasma with the photospheric magnetic field, which results in the confined plasma rotating at velocities that increase linearly with distance from the star. Beyond the Kepler corotation radius \rk, centrifugal support is stronger than gravity and gravitational infall of the confined plasma is prevented \citep[e.g.][]{ud2008}. This leads to an accumulation of material above \rk, and a cavity below \rk, thereby giving rise to the double-humped morphology \citep[e.g.][]{town2005c}. A Keplerian disk seen edge-on can also produce a double-humped emission profile, however the crucial difference with $\sigma$ Ori E-type emission is that the emission of a Keplerian disk is confined inside $\pm$\vsini.

To analyze the time variability of H$\alpha$ we measured its EWs. We utilized the individual Stokes $I$ spectra in order to maximize the size and time-resolution of the dataset (i.e.\ 108 spectra). EWs were measured in the red half of the line, from $+$\vsini~to the red edge of emission at $+700$~\kms, and in the blue half of the line between the blue edge of emission at $-700$~\kms~and $-$\vsini. The core of the line was excluded so as to avoid contaminating the measurements with nebular emission (Fig.\ \ref{W601_halpha_minmax}). To correct for the EW variation due to the radial velocity variations of the two components, for each spectrum we calculated synthetic binary spectra in the same fashion as was done for H$\beta$ in determining $\log{g}$ (Fig.\ \ref{w601_bin_logg_fit}), using the best-fit parameters for each star and a radius ratio of $R_{\rm A}/R_{\rm B} = 1.1$ as inferred from spectroscopic modelling (Fig.\ \ref{tlusty_fit}). The EWs of the model spectra were measured within the same integration ranges, and subtracted from the EWs of the data. These EWs are used to determine the rotational period below in \S~\ref{subsec:period_determination}. Line profile variations are examined in greater detail in the context of the magnetospheric analysis in \S~\ref{subsec:breakout}.

Magnetic hot stars typically have surface chemical abundance spots that lead to rotational modulation of EW measurements from photospheric line profiles. We examined the He~{\sc i}~667.8~nm, Si~{\sc iii}~455.4~nm, and C~{\sc ii}~426.7~nm lines, choosing these lines on the basis of being relatively strong and isolated. The EWs of these line were measured from individual Stokes $I$ subexposures across the full line width, as inferred by eye, with the individual spectra renormalized using a linear fit to adjacent continuum regions.  There is no statistically significant variability in either Si~{\sc iii} or C~{\sc ii} (the reduced $\chi^2$ for the null assumption of no variation about the mean value being in both cases about unity). He~{\sc i}, however, exhibits statistically significant variation, and this line was also included in the period analysis in \S~\ref{subsec:period_determination}.

\section{Rotation}\label{sec:rot_em}

\subsection{Period determination}\label{subsec:period_determination}

   \begin{figure}
   \centering
   \includegraphics[width=0.45\textwidth]{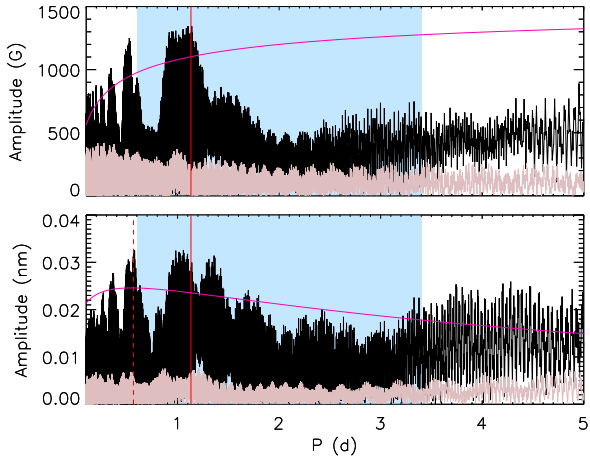} 
      \caption[]{Periodograms for \bz~(top) and H$\alpha$ EWs (bottom). Shaded regions indicate the physically plausible range of periods (see text). Periodograms are shown in black. The 3$\sigma$ noise level is shown by purple curves. The adopted rotation period is indicated with a solid red line. For the H$\alpha$ periodogram, red dashed lines indicate rotational harmonics. Grey periodograms are after pre-whitening with the rotation period (top) or the rotation period and its harmonics (bottom).}
         \label{w601_bz_ew_periods}
   \end{figure}

   \begin{figure}
   \centering
   \includegraphics[width=0.45\textwidth]{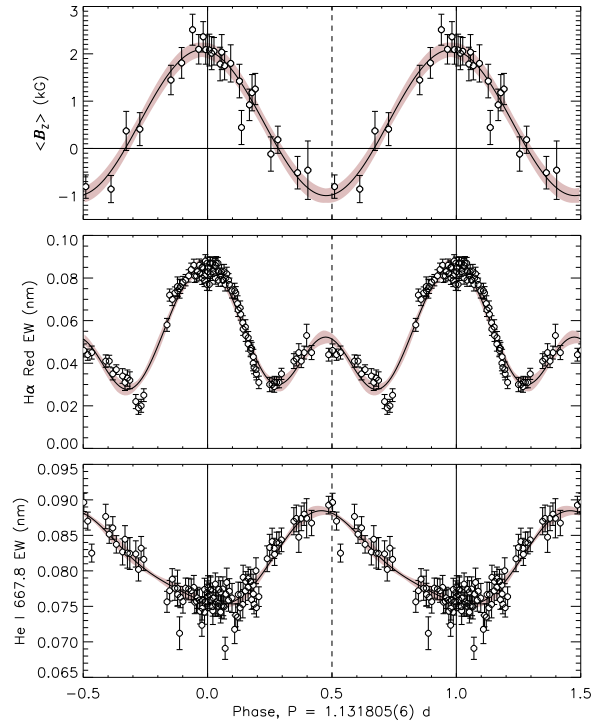} 
      \caption[]{\bz~(top), red H$\alpha$ emission EW (middle), and He~{\sc i}~667.8~nm EW (bottom), phased with the rotation period. Curves and grey shaded regions show harmonic fits to the data and 1$\sigma$ uncertainties. Solid and dashed vertical lines indicate magnetic extrema. EWs were obtained from individual Stokes $I$ sub-exposures. All data have been sigma-clipped to remove the noisiest observations. H$\alpha$ emission peaks occur at the extrema of \bz, as expected for rotationally modulated magnetospheric emission.}
         \label{w601_halpha_bz}
   \end{figure}

Magnetic hot stars generally exhibit spectropolarimetric and spectroscopic variations modulated precisely by the rotation period, with no inter-cycle variations i.e.\ the variability is perfectly regular. Therefore, the rotation period can be determined independently using multiple diagnostics. The period analysis of individual diagnostics (\bz, H$\alpha$ and He~{\sc i}~667.8 nm EWs) is described in the following, with 1.131805(6)~d from H$\alpha$ EWs being adopted as the most likely rotational period.

As a first attempt to determine the rotation period the \bz~measurements were analyzed using {\sc period04}. The resulting periodogram is shown in the top panel of Fig.\ \ref{w601_bz_ew_periods}. The shaded region indicates the range of physically plausible rotation periods, with the lower bound set by the breakup velocity and the upper bound set by $R_*$ and \vsini. Within this window, the highest peak is at 1.13980(3)~d (with the uncertainty in the least significant digit given in parentheses), with a $S/N$ of 9, i.e.\ the period is above the threshold of 4 usually adopted for formal significance \citep{1993A&A...271..482B,1997A&A...328..544K}. However, there are numerous other nearby peaks with a similar amplitude, and it is not possible to distinguish between them based purely on the coherence of the phase variation.

We also analyzed the EWs described in \S~\ref{sec:halpha_em} using {\sc period04}. The H$\alpha$ periodogram is shown in the bottom panel of Fig.\ \ref{w601_bz_ew_periods}. There is a single strong peak at about 1.15 d, consistent with the forest of peaks obtained for \bz. However, the strongest signal is at 0.56589 d, very close to the first harmonic of the period identified in \bz. This is typical of the double-wave EW variations produced by the CMs of stars in which both magnetic poles are visible during a rotation cycle, as is the case with W\,601\,B \citep[e.g.][]{lb1978,2020MNRAS.499.5379S}. Under the presumption that this is one-half of the true rotation period, we fit the period and this harmonic with {\sc period04} in order to obtain a period of 1.13179(1)~d. The $S/N$ of the rotational period and its first harmonic are respectively 7 and 12. If EWs measured from the red half of the line (\S~\ref{sec:halpha_em}) are analyzed in isolation -- where the emission strength is at a maximum and the amplitude of variation is the largest -- the respective $S/N$ of the rotation period and its first harmonic are 27 and 29, with $P_{\rm rot} = 1.131805(6)$~d. Analyzing the blue H$\alpha$ EWs in isolation does not yield useful results. The strongest peak in the periodogram is at about 0.558783(6)~d, qualitatively consistent although formally inconsistent with the first harmonic of the period obtained from the red EWs; however, the $S/N$ of this period is not high (5.6). We adopt the period obtained from the red EWs as the most precise, and the most likely to be correct.

For He~{\sc i}~667.8~nm, the strongest peak is at 1.09099(2) d. However, if the He~{\sc i} EWs are phased with the H$\alpha$ period, {\sc period04} returns a higher amplitude (0.006 nm vs.\ 0.005 nm) and a higher $S/N$ (11.5 vs.\ 9.8). Since H$\alpha$ yields a higher $S/N$, the slightly different period returned by He~{\sc i}~667.8~nm is probably a consequence of the lower $S/N$ of the variation of this line (the semi-amplitude of variation in this line being less than 3$\times$ the mean error bar).

\subsection{Rotational Modulation}\label{subsec:rotmod}

\bz, H$\alpha$ EWs, and He~{\sc i}~667.8~nm EWs are shown phased with $P_{\rm rot}$ in Fig.\ \ref{w601_halpha_bz}, using $JD0 = 2453957.4(2)$ as determined by fitting a sinusoid to \bz~and determining the \bz~maximum one cycle before the first observation. The reduced $\chi^2$ of a first-order sinusoidal fit to \bz~is 1.3, indicating that a sinusoid is a good fit and that \bz~can therefore be reproduced by a simple tilted dipole, as is the case for most magnetic early-type stars \citep[e.g.][]{2018MNRAS.475.5144S,2019A&A...621A..47K}. 

The He~{\sc i} EWs phase coherently with the rotational period (Fig.\ \ref{w601_halpha_bz}, bottom), with the extrema of the He~{\sc i} EW curve corresponding to the \bz~extrema. The He variability pattern suggests that the strongest He abundance spot is at the negative magnetic pole. Coherent EW variation is notable in the case of a PMS star, as it suggests that surface chemical abundance patches form almost immediately in the photospheres of magnetic hot stars.

The rotational modulation of H$\alpha$ is examined in detail in \S~\ref{subsec:breakout}.

   \begin{figure*}
   \centering
   \includegraphics[width=0.95\textwidth, trim = 100 50 100 50]{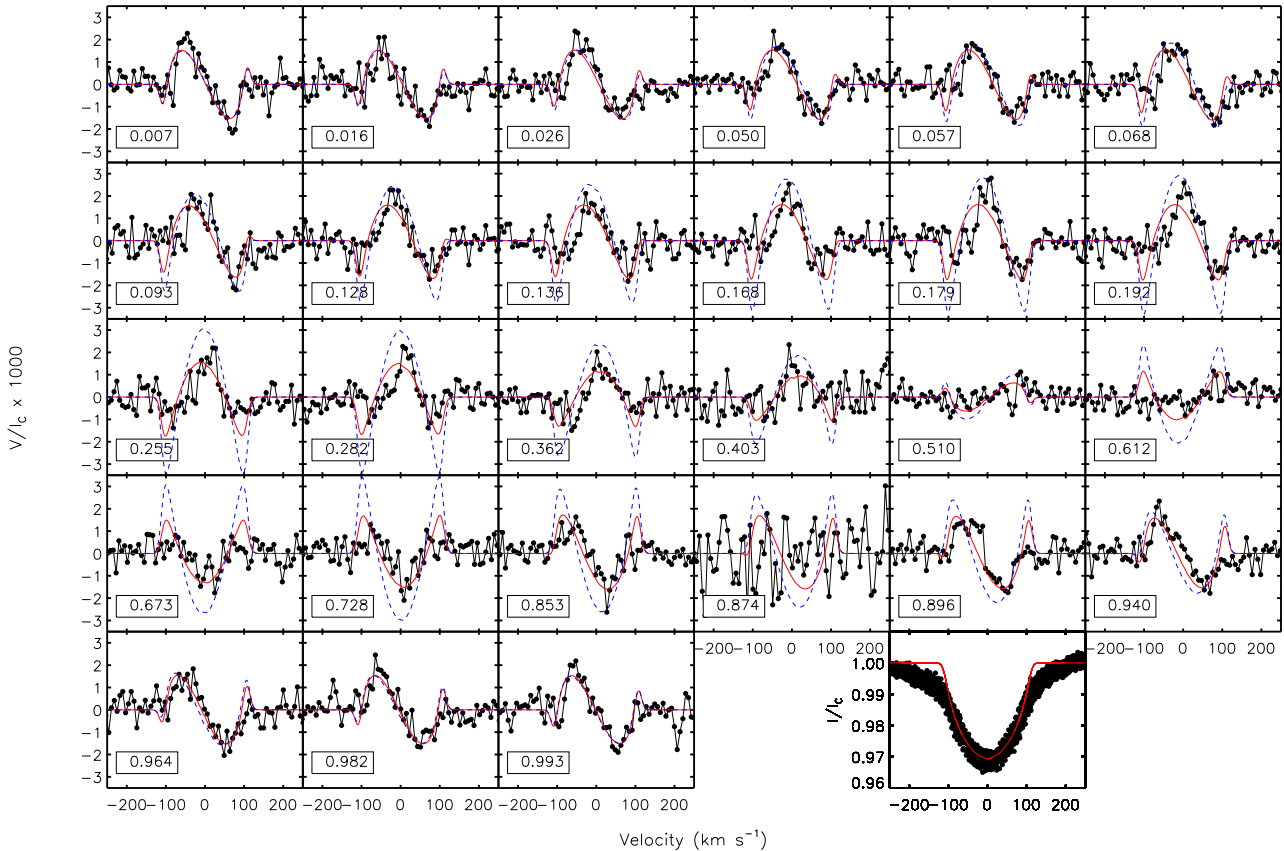} 
      \caption[]{Observed LSD Stokes $V$ profiles (black circles) compared to synthetic Stokes $V$ profiles inferred from the \bz~curve model (dashed blue) and from direct Bayesian modelling of Stokes $V$ (solid red). Rotational phases are indicated in boxes in the upper left corner of each panel. The bottom right panel shows the overplotted distenangled Stokes $I$ profiles of W\,601\,B, as compared to the synthetic Stokes $I$ profile (red line) used to model Stokes $V$. Neither model provides a fully satisfactorily reproduction of the Stokes $V$ variation.}
         \label{lsd_fit}
   \end{figure*}

\section{Magnetic modelling}\label{sec:orm}

\begin{table}
\centering
\caption[]{Rotational, magnetic, and magnetospheric parameters for W\,601\,B (see text for definitions.)}
\label{magtab}
\begin{tabular}{l c }
\hline\hline
Parameter & Value  \\
\hline
$P_{\rm rot}~({\rm d})$ & $1.13178 \pm 0.00001$ \\
$T_0~({\rm HJD})$ & $2453957.4 \pm 0.2$ \\
$i_{\rm rot}~(^\circ)$ & $31^{+5}_{-3}$ \\
$v_{\rm eq}~({\rm kms})$ & $161^{+33}_{-9}$ \\
$W$ & $0.24^{+0.09}_{-0.01}$ \\
$R_{\rm p}/R_{\rm e}$ & $0.971^{+0.003}_{-0.03}$ \\
$R_{\rm K}~({\rm R_*})$ & $2.1^{+0.2}_{-0.2}$ \\
\hline
$B_0~({\rm kG})$ & $0.54 \pm 0.07$ \\
$B_1~({\rm kG})$ & $1.55 \pm 0.09$ \\
$\beta~(^\circ)$ & $79^{+1}_{-3}$ \\
$B_{\rm d}~({\rm kG})$ & $11^{+3}_{-1}$ \\
\hline
$\log{(\dot{M} / {\rm M_\odot / yr})}$ & $-9.5 \pm 0.1$ \\
$v_\infty~({\rm km/s})$ & $1120 \pm 60$ \\
$\log{\eta_*}$ & $6.0 \pm 0.2$ \\
$R_{\rm A}~({\rm R_*})$ & $31^{+5}_{-2}$ \\
$\log{R_{\rm A}/R_{\rm K}}$ & $1.2 \pm 0.1$ \\
$\log{(B_{\rm K} / {\rm G})}$ & $2.7 \pm 0.2$ \\
$\log{(\tau_{\rm J} / {\rm yr}})$ & $6^{+0.3}_{-0.1}$ \\
$\log{(t_{\rm S,max} / {\rm yr}})$ & $5.8^{+0.5}_{-0.1}$ \\
\hline\hline
\end{tabular}
\end{table}

The rotationally magnetic magnetic variability of hot stars is described using the Oblique Rotator Model \citep[ORM; e.g.][]{1950MNRAS.110..395S}, in which the sinusoidal variation in \bz~of a dipolar magnetic field rotating in the plane of the sky is parameterized with the inclination $i_{\rm rot}$ of the rotational axis from the line of sight, the tilt angle $\beta$ of the magnetic axis from the rotational axis, and the strength of the magnetic dipole at the stellar surface $B_{\rm d}$.

To determine W\,601\,B's rotational, ORM, and magnetospheric parameters, we utilized the Monte Carlo (MC) Hertzsprung-Russell Diagram (HRD) sampler described by \cite{2019MNRAS.490..274S}. The MC sampler combines information about a star's observed atmospheric, magnetic, and rotational properties, together with ancillary information such as e.g.\ the age of its parent cluster, with evolutionary models in order to infer fundamental stellar parameters, Oblique Rotator Model (ORM) parameters, rotational parameters, and magnetospheric parameters. These are given in Table \ref{magtab}. 

As inputs we used the stellar parameters obtained above (Fig.\ \ref{physpar}, Table \ref{partab}), the star's rotational period, and \vsini~to obtain $i_{\rm rot} = 31^{\circ+5}_{-3}$. This is similar to the orbital axis inclination $i_{\rm orb} = 31.5 \pm 0.9^\circ$, i.e.\ the spin and orbital axes of the system are aligned or nearly aligned. The sinusoidal fitting parameters to \bz~$=B_0 + B_1\sin{(\phi + \Phi)}$ are $B_0 = 540 \pm 70$~G, $B_1 = 1550 \pm 90$~G, and phase offset $\Phi = 1.67 \pm 0.09$~rad (from the model shown in the top panel of Fig.\ \ref{w601_halpha_bz}). From the geometrical relations given by \cite{preston1967,preston1974}, and with the maximum measured value of \bz~being \bz$_{\rm max} = 2509 \pm 327$~G, the magnetic axis obliquity angle is then $\beta = 79^{\circ+1}_{-3}$ and the surface magnetic dipole strength is $B_{\rm d} = 11^{+3}_{-1}$~kG \citep[using a linear limb darkening coefficient $\epsilon = 0.4$ from line-blanketed NLTE model spectra;][]{2016MNRAS.456.1294R}. 

Another means of determining W\,601\,B's ORM parameters is via direct modelling of Stokes $V$ using a version of the Bayesian inference method described by \cite{petit2012a} modified to include rotational phase information. The results of this fit are shown compared to observations in Fig.\ \ref{lsd_fit}. Direct modelling of Stokes $V$ produces a best fit for $i_{\rm rot}=54^{+20^\circ}_{-5}$, $\beta=60^{+5^\circ}_{-10}$, and $B_{\rm d} = 6.2^{+0.8}_{-0.4}$~kG, where the uncertainties correspond to the 68.7\% credible regions. These values differ substantially from the values inferred from \bz~and the stellar parameters, although values of $B_{\rm d}$ comparable to those inferred from \bz~can be accommodated within the 95.4\% credible region. Note that, unlike \bz~fitting, direct modelling of the Stokes $V$ profile does not constrain $i_{\rm rot}$ from $R_*$, \vsini, and $P_{\rm rot}$. The smaller value of $B_{\rm d}$ found by modelling Stokes $V$ is due to the larger value of $i_{\rm rot}$: $i_{\rm rot} < 45^\circ$ is excluded by the Stokes $V$ fits, whereas $i_{\rm rot} > 30^\circ$ is excluded by the star's radius and rotational properties. The Stokes $V$ model and the \bz~model agree well near magnetic maximum at phase 0. However, the \bz~model predicts a crossover signature with a larger amplitude as the magnetic equator comes into view near phases 0.2 and 0.7. A possible reason for this discrepancy may be that, notwithstanding the reasonable fit of a dipole model to \bz, the magnetic field is not purely dipolar but instead a `distorted dipole' \citep[the most common toplogy revealed by Zeeman Doppler Imaging;][]{2019A&A...621A..47K}. Supporting this supposition, neither set of ORM parameters provides a faithful reproduction of Stokes $V$ at all phases.

An alternative explanation for the tension between the ORM parameters inferred from \bz~and those obtained via direct modelling of Stokes $V$ is systematic error in W\,601\,B's stellar parameters. If W\,601\,B is actually at the ZAMS, then assuming a mass of $5.7 \pm 0.3$~\msun~its minimum possible radius is $R_* = 2.9 \pm 0.1$. This results in $i = 44 \pm 3^\circ$, $\beta = 75 \pm 3^\circ$, and $B_{\rm d} = 7.9^{+1}_{-0.7}$~kG, very similar to the values found via modelling of Stokes $V$. However, this would require $\log{g} = 4.3$, in which case W\,601\,A would need to have a surface gravity of $3.7$ (left panel of Fig.\ \ref{w601_bin_logg_fit}). In this case, the age of the system would need to be about $\log{t} = 7.2$, which is much older than the NGC\,6611 cluster ($\log{t} \sim 6.75$). W\,601\,A would also have a much larger radius than W\,601\,B, which is inconsistent with spectroscopic modelling, which instead indicate their radii are almost identical.

\subsection{Magnetic Constraints on the Primary}\label{subsec:primmag}

The line profiles of the two stellar components are blended in all observations, making it difficult to constrain the magnetic properties of W\,601\,A. To do this, we subtracted the model fits to Stokes $V$ obtained via Bayesian inference (Fig.\ \ref{lsd_fit}) from the observed Stokes $V$ profiles, yielding composite LSD profiles consisting of the disentangled Stokes $I$ profiles of W\,601\,A and the residual Stokes $V$ profiles. We measured FAPs and \bz~in the usual fashion, with an integration range of $\pm 220$~\kms around the rest frame of W\,601\,A. All of the profiles are non-detections, indicating that, despite the imperfections in the fit to Stokes $V$, the residuals are not consistent with statistically significant departures from null polarization. The mean \bz~error bar is 450 G, and the root-mean-square \bz~is 365 G, further confirming that the Zeeman signature was successfully removed.

We then analyzed the profiles using the same Bayesian inference engine used to infer ORM parameters for W\,601\,B, with the difference that, since $P_{\rm rot}$ is unknown, rotational phase was treated as a nuisance parameter \citep{petit2012a}. The Stokes $V$ profiles yielded upper limits on $B_{\rm d}$ at 68.3\%, 95.4\%, 99.0\%, and 99.7\% credible regions of 306 G, 826 G, 1704 G, and 2832 G respectively. The corresponding upper limits obtained from $N$ are comparable, respectively 279 G, 779 G, 1590 G, and 2640 G. 

Any possible magnetic field possessed by W\,601\,A is less than W\,601\,B's by a factor of at least 2 with 99.7\% credibilty. With 68.3\% credibility its magnetic field is less than 300 G, which is the approximate critical magnetic field strength necessary for the magnetic field to maintain stability against rotational or convective instabilities \citep{2007A&A...475.1053A,2019MNRAS.487.3904M,2020ApJ...900..113J}. Fossil magnetic fields weaker than 300 G are exceptionally uncommon \citep{2007A&A...475.1053A,2019MNRAS.483.3127S,2019MNRAS.490..274S}, and those magnetic fields that have been detected below this threshold are usually either in evolved stars \citep[e.g.][]{2015AA...574A..20F,2017MNRAS.471.1926N,2018MNRAS.475.1521M} or ultra-weak fields (on the order of $\sim 0.1 - 10$~G) as found in Vega \citep{2009AA...500L..41L,2010AA...523A..41P}, or in Am stars such as Sirius \citep{2011AA...532L..13P}, Alhena \citep{2016MNRAS.459L..81B,2020MNRAS.492.5794B}, and others. While no ultra-weak field has yet been detected in a B-type star, the similarly of the magnetic properties of A and B-type stars \citep{2019MNRAS.490..274S} suggests that the `magnetic desert' between the ultra-weak fields and fossil magnetic fields likely persists across the entire upper main sequence. Since the upper limit on W\,601\,A's magnetic field is comparable to the critical magnetic field strength, any potential magnetic field is likely to be on the order of a few tens of G or less. 

\section{Discussion}\label{sec:discussion}

\subsection{Magnetosphere}\label{subsec:breakout}

   \begin{figure}
   \centering
   \includegraphics[trim=50 50 50 50, width=.45\textwidth]{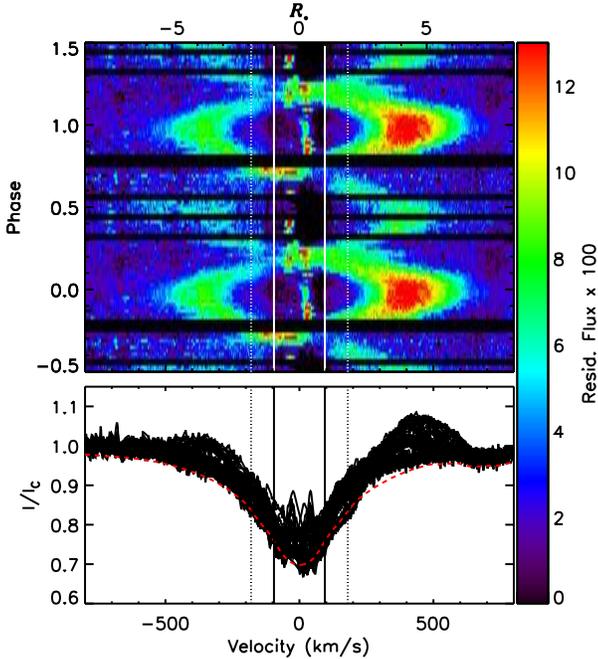} 
      \caption[]{H$\alpha$ dynamic spectrum. The top panel shows residual flux, shifted to the rest frame of W\,601\,B, folded with the rotational period, with intensity corresponding to the colour bar. The bottom panel shows the observed spectra (black) and an example synthetic spectrum (red). Residual flux was determined by subtracting synthetic binary spectra tailored to each observation. The vertical solid and dashed lines show $\pm v\sin{i} = \pm R_*$, and $\pm R_{\rm K}$, for W\,601\,B. The asymmetry in emission strength between red and blue emission bumps at phase 0 is likely indicative of a departure from a pure dipole. There is additionally no indication of eclipses in the core of the line.}
         \label{W601_halpha}
   \end{figure}

Rapidly rotating, strongly magnetic B-type stars frequently display $\sigma$ Ori E variability originating in circumstellar magnetospheres \citep{lb1978}. H$\alpha$ (Fig.\ \ref{w601_halpha_bz}, middle) shows a double-wave variation, with the strongest of the maxima corresponding to the \bz~maximum and the weaker local maximum corresponding to the \bz~minimum (vertical lines in Fig.\ \ref{w601_halpha_bz}). This double-wave EW variation is consistent with line formation within a Centrifugal Magnetosphere (CM) formed in a tilted dipole \citep{petit2013}. In such a case the CM is expected to be shaped like a warped disk, with the two densest regions at the intersections of the rotational and magnetic equators \citep[e.g.][]{town2005c}. When the magnetic pole is closest to the line of sight (i.e.\ at maximum $|\langle B_z \rangle|$), the projected area of the CM, and therefore the EW of H$\alpha$, is at a maximum. If the second magnetic pole is visible at some point during the rotation cycle, this should correspond to a secondary emission peak. This pattern is indeed observed in most CM host stars, the exceptions being those stars with very complex surface magnetic fields \citep{2020MNRAS.499.5379S}.

From the star's fundamental and atmospheric parameters, the \cite{vink2001} mass-loss recipe yields a mass-loss rate of $\log{\dot{M}} = -9.5 \pm 0.1~{\rm M_\odot / yr}$ and a wind terminal velocity of about $1000$~\kms. From Eqn.\ 7 in \cite{ud2002}, the wind magnetic confinement parameter (that is, the ratio of magnetic to kinetic energy density at the magnetic equator at the stellar surface) is $\log{\eta_*} = 6.0 \pm 0.2$; since $\eta_*$ is greater than unity the wind is magnetically confined. From Eqn.\ 9 in \cite{ud2008}, $\eta_*$ scales to an Alfv\'en radius (i.e.\ the maximum extent of magnetic confinement) of \ra~$=32 \pm 4 R_*$. 

The Lorentz force enforces corotation of magnetically confined plasma with the stellar magnetic field out to \ra. A CM forms when the Kepler corotation radius \rk, defined as the distance at which centrifugal and gravitational forces are balanced, is less than \ra. The Kepler radius is obtained from the critical rotation parameter $W = v_{\rm orb}/v_{\rm eq} = 0.24^{+0.1}_{-0.01}$ via the scaling $R_{\rm K}/R_* = W^{-2/3}$, where $v_{\rm orb} = \sqrt{G M_*/R_*}$ is the orbital velocity i.e.\ the velocity required for a Keplerian orbit at the stellar equator \citep{ud2008}. The equatorial rotational velocity is $v_{\rm eq} = 161^{+32}_{-9}$~\kms, with a moderate oblateness ratio of the polar to equatorial radii of $R_{\rm p}/R_{\rm e} = 0.97 \pm 0.02$. The Kepler corotation radius is then $R_{\rm K} = 2.1 \pm 0.2~R_*$ \cite[using Eqn.\ 14 from][]{ud2008}. This is a similar Kepler radius to those seen for other H$\alpha$-bright CM host stars \citep{2019MNRAS.490..274S}. \rk~is indicated in Fig.\ \ref{W601_halpha_minmax}: as expected, the strongest emission is above the Kepler radius. The ratio of \ra~to \rk~is \rark~$=1.2 \pm 0.1$: not only is \ra~significantly greater than \rk, but \rark~is well within the typical range for H$\alpha$-bright CM host stars \citep{2019MNRAS.490..274S}. 

A dynamic spectrum of W\,601\,B's H$\alpha$ line is shown in Fig.\ \ref{W601_halpha}. This was created by subtracting the synthetic spectra used to correct the EWs for the RV variation of the two stars (\S~\ref{sec:halpha_em}, Fig.\ \ref{W601_halpha_minmax}) from the observed spectra, with the residual flux shifted to the rest frame of W\,601\,B. Due to the rigid rotation of the CM, there is a linear relationship between the projected distance from the star and velocity, i.e.\ $r/R_* = v_{\rm r}/v\sin{i}$ (where $v_{\rm r}$ is the line-of-sight velocity), as indicated on the top horizontal axis. At maximum emission, there are two emission bumps extending out to several stellar radii, with the strongest emission at about 3.8~$R_*$. These emission bumps move closer to the line centre as they weaken in strength, corresponding to the reduction of the projected areas of the clouds simultaneous with the reduction in their projected distance from the star. The secondary emission maximum at phase 0.5 has a similar extent in velocity space, but lower peak emission strength, indicating that the CM is more flattened in projection at this phase, which is consistent with the negative pole not coming as close to alignment with the line of sight. 

There is stronger emission in the rotationally broadened core of the line around phases 0.25 and 0.75 than is present in this part of the line at other phases. There is nebular emission contaminating the very centre of the line (see also Fig.\ \ref{W601_halpha_minmax}), however this nebular emission is much narrower than signatures associated with the CM near phases 0.25 and 0.75. There are no enhanced absorption features in the line core at these phases, and therefore no evidence that the star is eclipsed by its CM \citep[as seen for $\sigma$ Ori, HR\,7355, or HD\,176582;][]{oks2012,rivi2013,bohl2011}. The presence or absence of eclipses is highly dependent on the magnetic geometry. No eclipses are expected if $i_{\rm rot} \le 35^\circ$ \citep{town2008}, as inferred from \bz. On the other hand, the ORM parameters inferred from direct modelling of Stokes $V$ predict two prominent eclipses near phase 0.5 \citep{town2008}.

As noted in \S~\ref{sec:orm}, there is some suggestion that W\,601\,B's surface magnetic field is not purely dipolar. The strong red-blue asymmetry in the H$\alpha$ emission, especially prominent near phase 0.0 (Fig.\ \ref{W601_halpha}) can only be explained if the surface magnetic field is not a pure dipole. 

\cite{2020MNRAS.499.5379S} and \cite{2020MNRAS.499.5366O} showed that the peak emission strength of stars with CMs is governed by centrifugal breakout, a magnetic reconnection mechanism whereby the magnetic field is overloaded by the wind-fed plasma and is explosively ejected away from the star \citep{ud2006,ud2008}. The greater the strength $B_{\rm K}$ of the magnetic field at the Kepler radius, the greater the density of plasma that the magnetosphere can hold, the more extensive the optically thick part of the H$\alpha$-emitting CM, and the stronger the emission. W\,601\,B's maximum emission strength is about 0.15 nm. W\,601\,B contributes about 33\% of the total light at H$\alpha$ (as determined from the two stars' radii and synthetic BSTAR2006 spectra for their respective atmospheric parameters), and this emission strength should therefore be scaled up by a factor of about 3, to a single-star emission strength of 0.45 nm. The strength of the equatorial magnetic field at W\,601\,B's Kepler radius is $\log{B_{\rm K}} = 2.7 \pm 0.2$, using the dipole strength inferred from \bz. Utilizing the centrifugal breakout emission strength scaling law developed by \cite{2020MNRAS.499.5366O}, a star with this value of $B_{\rm K}$ and W\,601\,B's stellar and rotational parameters has an expected emission strength of $0.4 \pm 0.1$ nm, almost exactly the measured value. Notably, this emission strength puts W\,601\,B in the company of the CM host stars with the strongest H$\alpha$ emission \citep[namely $\sigma$ Ori E and HD\,345439;][]{2020MNRAS.499.5379S}. Using the dipole strength inferred from Stokes $V$ yields $\log{B_{\rm K}} = 2.65^{+0.13}_{-0.22}$ and an inferred emission equivalent width of about 0.2 to 0.3 nm, still amongst the strongest known.

The peak emission strength of W\,601\,B's CM occurs at about 3.8~$R_*$, which is about 1.8~$R_{\rm K}$. \cite{2020MNRAS.499.5379S} noted that \rk~is systematically less than the radius of maximum emission, and the discrepancy between the radius of maximum emission and \rk~is in this case at the top of the range found for their sample (see their Fig.\ 12). As discussed above in \S~\ref{sec:orm}, the conflict between ORM parameters inferred via modelling of \bz~and via modelling of Stokes $V$ can be reconciled if the radius is smaller than was assumed when deriving ORM parameters from \bz. This smaller radius also yields a larger Kepler radius, $R_{\rm K} = 2.80 \pm 0.03~R_*$. This is still only about 70\% of the radius of emission maximum, therefore adopting the minimum possible radius cannot resolve the discrepancy in the case of this star.

Binarity is unlikely to have any effect on W\,601\,B's H$\alpha$ magnetosphere. The periastron separation of the stars is about 0.9 AU (see Tables \ref{orbtab} and \ref{partab}), whereas the Alfv\'en radius of W\,601\,B is a maximum of 37 $R_*$ or, with $R_* = 4.4$~\rsun, about 0.75 AU. W\,601\,A is thus outside of W\,601\,B's magnetosphere at all orbital phases. This argument is of course sensitive to the mass-loss rate prescription: the much lower \cite{krticka2014} mass-loss rates imply $R_{\rm A} = 62^{+22}_{-6}~R_*$, i.e.\ about $1.3^{+0.4}_{-0.1}$ AU, in which case W\,601\,A is inside the magnetosphere at essentially all orbital phases. In either case, however, W\,601\,A is very far from the distance of H$\alpha$ line formation, a maximum of $7 R_* = 0.15$~AU from W\,601\,B. 

It is interesting to note that W\,601 emits gyrosynchrotron radiation \citep{2017MNRAS.465.2160K}, which is believed to be generated by electrons accelerated to relativistic energies within the middle magnetosphere current sheet just outside the Alfv\'en radius \citep{2004A&A...418..593T}. If W\,601\,B's current sheet extends far enough beyond \ra~\citep[which is by no means excluded by the models developed by][]{2004A&A...418..593T}, it may be possible that W\,601\,A disrupts the current sheet during periastron passage. It would be of interest to obtain radio observations of this system at apastron and periastron, in order to look for such an effect.

\subsection{Rotational and Magnetic Evolution}\label{subsec:rotev}

Amongst the magnetic high-mass pre-main sequence stars, W\,601\,B has both the strongest magnetic field and one of the shortest rotation periods. Of the magnetic PMS stars noted in the introduction, the mean surface dipole strength is about 1.2 kG, and the strongest belongs to V\,380\,Ori \citep[with a strength of 2.2 kG;][]{2009MNRAS.400..354A}, while the measured rotation periods of other magnetic PMS hot stars range from 4 d to 40 d \citep{2009MNRAS.400..354A,2015A&A...584A..15J}. This makes W\,601\,B's rotational and magnetic evolution of some interest. 

Main sequence magnetic hot stars experience simultaneous decline in surface magnetic field strength, due to a combination of magnetic flux conservation in an expanding atmosphere and magnetic flux decay, and rapid magnetic braking due to angular momentum loss \citep{ud2008,2019MNRAS.490..274S}. As a PMS star, W\,601\,B is still contracting towards the ZAMS. Therefore, flux conservation would imply that its surface magnetic field should grow stronger as the radius shrinks, while at the same time conservation of angular momentum would normally mean that the star should spin up. In this case, the current radius of about 4.4~\rsun~should shrink to about 3.2 \rsun~at the ZAMS. The total unsigned magnetic flux is $\Phi = B_{\rm d} R_*^2$, therefore if flux is conserved then the ZAMS field should increase to about $21^{+5}_{-2}$ kG. This is comparable to the top of the range of surface magnetic dipole strengths of magnetic early B-type stars seen close to the ZAMS \citep{2019MNRAS.490..274S}. 

At the same time, conservation of angular momentum means that the star should spin up as it approaches the ZAMS. Under the assumption that angular momentum is conserved, the ZAMS equatorial rotational velocity should be $v_{\rm eq, ZAMS} = v_{\rm eq} R_* / R_{\rm ZAMS} = 250^{+46}_{-15}$~\kms, corresponding to a rotational period $P_{\rm rot,ZAMS} \sim 0.7$~d. This is similar to the two most rapidly rotating magnetic B-type stars known, HR\,5907 and HR\,7355 \citep{grun2012,rivi2013}, which have periods of about 0.5 d.

Angular momentum of course should not be conserved, since it is being lost via the magnetosphere. From Eqn.\ 20 in \cite{ud2009}, the spindown timescale $\tau_{\rm J}$ -- i.e.\ the e-folding timescale for the rotation parameter $W$ -- is $\log{(\tau_{\rm J} / {\rm yr})} = 5.9^{+0.3}_{-0.1}$ or between 700 kyr and 1.5 Myr, using the \cite{vink2001} mass-loss rates. If \cite{krticka2014} mass-loss rates are used instead, $\log{(\tau_{\rm J} / {\rm yr})} = 6.5^{+0.3}_{-0.1}$. Given the current age of the star it should reach the ZAMS in about 300~kyr, i.e.\ about half an e-folding timescale using Vink mass-loss, and much shorter than $\tau_{\rm J}$ using \citeauthor{krticka2014} mass-loss. As the star evolves towards the ZAMS, its rotational evolution should therefore be dominated by spin-up due to contraction.

We conducted an analysis of the rotational evolution of W\,601\,B using the 7 \msun~PMS \cite{Haemmerle2019} evolutionary track, simultaneously accounting for angular momentum loss via the magnetosphere, spin-up due to contraction towards the ZAMS, and the change in $B_{\rm d}$ due to flux conservation. The model had an initial dipole strength $B_{\rm d,init} = 11$~kG. We started the model at 4~\rsun~(corresponding to an age of 400 kyr from the birthline), with $W = 0.3$ (corresponding to $P_{\rm rot} \sim 1.13$~d). The minimum rotation period in this case is achieved at 490 kyr, with $P_{\rm rot} \sim 0.55$~d; after this point magnetic braking dominates the rotational evolution. This corresponds to a spin-up of $-0.5$~s/yr. More accurate predictions of the spin-up rate may be provided by the development of a PMS extension of self-consistent stellar evolutionary models incporating fossil fields, similar to those presented by \cite{2020MNRAS.493..518K}. 

Rotational evolution has been detected in several magnetic early-type stars. $\sigma$ Ori E is spinning down at a rate of about $+0.08$~s/yr, consistent with expectations from magnetic braking \citep{town2010}. HD\,37776 and CU\,Vir both exhibit complex patterns of cyclical spin-up and spin-down, \citep{miku2008,miku2011,2017ASPC..510..220M}, with typical period changes on the order of 0.6 s/yr and 0.1 s/yr, respectively. HD\,142990 is spinning up at a rate of about $-0.6$~s/yr \citep{2019MNRAS.486.5558S}. It is intriguing to note that both HD\,37776 and HD\,142990 are very young stars \citep{2019MNRAS.490..274S}, and their spin-up rates are comparable to that expected for W\,601\,B. While contraction towards the ZAMS cannot explain the cyclical nature of HD\,37776's period evolution, it may provide an explanation for that of HD\,142990. In all cases in which rotational evolution has been directly measured, the datasets have spanned about 30 years; it is therefore not yet possible to detect rotational evolution in W\,601\,B, since the dataset for this star is extends across only 7 years, with the majority of the data having been acquired over a 2-year time span.

\cite{2019MNRAS.490..274S} noted that ultra-slow rotators such as $\xi^1$ CMa \citep{2017MNRAS.471.2286S,2018MNRAS.478L..39S} are difficult to explain using standard magnetic braking theory. In particular, they are so slowly rotating that their rotation periods are even longer than can be explained by standard magnetic braking theory under the usual assumption of initially critical rotation \citep{ud2009,petit2013}. One possible explanation is that such stars lose a great deal of angular momentum on the PMS. Indeed, their initial critical rotation fraction at the ZAMS would need to already be very close to 0. So far no such magnetic PMS B-type stars have been found \citep[although the magnetic Herbig Ae star HD\,101412 is quite a slow rotator, with a period of 42~d;][]{2015A&A...584A..15J}. Instead, W\,601\,B is apparently a precursor of stars that arrive at the ZAMS as rapid rotators. However, at this point relatively few magnetic PMS hot stars are known; identifying a larger sample is crucial to determining whether there is indeed a sub-population of PMS slow rotators that can serve as the progenitors for stars such as $\xi^1$ CMa, or whether some additional braking mechanism is necessary to explain ultra-slow rotation.

\subsection{Implications for the origin of fossil magnetic fields}

Using the same 7 \msun~evolutionary track, once again assuming magnetic flux conservation and evolving back in time from W\,601\,B's current magnetic parameters,  its surface magnetic dipole would have been at a minimum intensity of about 700 G approximately 140 kyr ago. Assuming the same $i_{\rm rot}$ and $\beta$ as presently, $B_{\rm d} = 700$~G would yield a maximum \bz~of about 130 G. This is comparable to the typical \bz~values observed by \cite{2019A&A...622A..72V} in their study of T-Tauri Stars (TTSs), Intermediate Mass T-Tauri Stars (IMTTSs), and Herbig Ae stars. 

This surface dipole strength is also comparable to the mean surface magnetic field strengths of TTSs and IMTTSs, which range from a few hundred G to a few kG \citep{2007MNRAS.380.1297D, 2008MNRAS.386.1234D, 2010MNRAS.409.1347D, 2011MNRAS.417.1747D, 2011MNRAS.417..472D, 2013MNRAS.436..881D, 2015MNRAS.453.3706D, 2009MNRAS.398..189H, 2015A&A...580A..39K, 2017MNRAS.472.1716H, 2017A&A...608A..77L, 2017MNRAS.467.1342Y, 2020MNRAS.497..632L, 2013ApJ...765...11J}. The continuity in total unsigned magnetic flux between stars at various stages on the PMS and MS, and across a wide range of masses, is suggestive of a common origin of fossil magnetic fields in dynamo processes occurring early in the evolutionary process. A striking property of several studies of magnetic binaries is that the magnetic properties of the two stars are often remarkably different even when their masses and rotational properties are nearly identical. For stars with convective envelopes this manifests as one star having a globally organized poloidal field and the other a more tangled magnetic topology. This is so for both V1878\,Ori, and the MS M dwarf visual binary GJ65 A and B \citep{2017ApJ...835L...4K}. In the case of the B9\,V `identical twin' eclipsing binary HD\,62658, one of the two stars is a chemically peculiar Bp star with a strong magnetic field, while the other is a normal, non-magnetic B-type star \citep{2019MNRAS.490.4154S}. Another example is provided by the B2\,IV binary HD\,149277, with a mass ratio of 1.1, in which one component is a He-strong star with a surface dipole strength of several kG, and the other has no detectable magnetic field \citep{2016PhDT.......390S,2018MNRAS.481L..30G}. While the mass ratio of W\,601 is much larger than these systems, it follows the general pattern of one component hosting a very strong magnetic field, while the upper limit for the magnetic field of the other is comparable to the 300 G critical field strength and therefore probably on the order of a few G (\S~\ref{subsec:primmag}), and confirms that this pattern is established already on the PMS.

The existence of several close binaries containing a magnetic star poses a challenge to the scenario in which dynamos generated during binary mergers are the primary pathway to fossil magnetic fields \citep[e.g.][]{2019Natur.574..211S}, since mergers should not produce close binaries. \cite{2019MNRAS.490.4154S} suggested that the properties of the binaries above point to an alternative scenario. During the convective stage of the star's evolution, irrespective of rotational properties or stellar mass, the dynamo magnetic field spontaneously settles into one of two attractor states: a globally organized dipole, or a tangled topology. Both leave behind a fossil field when the star becomes fully radiative. However, weaker or more tangled fields rapidly decay under the influence of rotational or convective instabilities \citep[e.g.][]{2007A&A...475.1053A,2019MNRAS.487.3904M,2020ApJ...900..113J}. In the case of close binaries, flux decay is accelerated via tidally induced instabilities \citep{2019A&A...629A.142V}. The result is that, very shortly after the cessation of convective support for a dynamo, only the strongest and most organized magnetic fields survive, while the incidence of magnetic fields amongst close binary stars is decreased even further by the additional influence of tidal instabilities. The abrupt change in magnetic incidence from essentially 100\% to the canonical 10\% seen for MS hot stars is precisely what is seen at the boundary between stars with convective envelopes and fully radiative stars \citep{2019A&A...622A..72V}.

\subsection{Formation of surface chemical abundance peculiarities}\label{subsec:diffusion}

The existence of He spots on the surface of W\,601\,B, as inferred from the rotational modulation of the He~{\sc i} 667.8 nm line (Fig.\ \ref{w601_halpha_bz}), indicates that surface chemical abundances become established quite rapidly on the PMS following the formation of the radiative envelope. It is instructive in this regard to compare W\,601\,B with other magnetic hot stars on the PMS. HD\,72106\,A (B9p) is the magnetic primary of a PMS SB2 system, with a non-magnetic Herbig Ae companion \citep{2008CoSka..38..245F}. The primary, which is either just reaching or has just passed the ZAMS, already possesses strong chemical peculiarities, while the non-magnetic companion has normal chemical abundances. This is similar to the case of W\,601. By contrast, the magnetic Herbig Ae star HD\,190073 has normal chemical abundances \citep{2007A&A...462..293C,folsom2012}, while another magnetic Herbig Ae star, HD\,101412 \citep{2005A&A...442L..31W,2009A&A...502..283H}, is underabundant in some elements \citep{2010A&A...523A..65C,folsom2012}. The abundances of HD\,101412 appear to reflect those of its dust-depleted accreting material \citep{2015A&A...582L..10K}. 

Notably, both HD\,190073 and HD\,101412 possess emission lines, which are most likely formed in accretion disks. HD\,72106\,A does not show Herbig emission, and while W\,601\,B is an H$\alpha$-bright star its emission originates in a CM and not an accretion disk. The scenario that emerges from this comparison is one in which magnetic hot stars that are still accreting possess surface abundances reflecting the accreting material. Once accretion ceases, diffusion in the magnetically stabilized radiative envelope leads to the rapid emergence of surface chemical peculiarities. Another interesting point of comparison is the PMS Am star Stock\,16\,12 \citep{2014MNRAS.442.3761N}. While Am stars do not possess strong magnetic fields, the existence of chemical peculiarities on a non-accreting PMS star is further evidence that these can form rapidly once accretion has ceased.

A possible contradiction to this scenario is V\,380\,Ori\,A, which shows both weak Bp-type chemical peculiarities \citep{folsom2012} and very strong Herbig-type H$\alpha$ emission. However, the secondary of this system is a chemically normal A-type star, which is much closer to the birthline, and it is very possible that the Herbig emission originates from this component. 

Atomic diffusion in PMS stars was evaluated by \cite{2011A&A...526A..37V} in order to determine how quickly the Ap/Bp phenomenon can be established. They found that, so long as turbulence is suppressed, surface chemical abundance peculiarities can form within 20 to 2 Myr, with the timescale decreasing with increasing stellar mass. While the minimum timescale is significantly longer than the PMS lifetime of W\,601\,B, the models utilized by \cite{2011A&A...526A..37V} extended only up to 2.8 \msun, much lower than W\,601\,B's mass. Although models of higher-mass PMS stars are needed to explore the parameter space occupied by W\,601\,B, the trend of decreasing timescales with increasing mass suggests that the formation of chemical spots on W\,601\,B's surface within a few hundred kyr is probably not in conflict with our current understanding. It is worth noting that suppression of turbulence within subsurface opacity-bump convection zones is both expected \citep{2019MNRAS.487.3904M,2020ApJ...900..113J} and observed \citep{2013MNRAS.433.2497S} in strongly magnetic early-type stars, and that the criterion that turbulence be suppressed is therefore very likely to be fulfilled.

\section{Conclusions}\label{sec:conclusion}

We have analyzed a large spectropolarimetric dataset of the magnetic pre-Main Sequence (PMS) star W\,601. The system turns out to be a spectroscopic binary with a mass-ratio of 1.8 and an orbital period of about 110 d. Atmospheric parameters (\teff~and $\log{g}$) were determined for the two stars via spectroscopic analysis. The \teff~of the primary, W\,601\,A, is about 22 kK, consistent with previous determinations for this star; the secondary, W\,601\,B, is slightly cooler at about 19 kK. W\,601\,A has a mass of about 12 \msun, and W\,601\,B a mass of about 7 \msun. 

Magnetic analysis reveals that the magnetic field belongs to W\,601\,B. The peak \bz~measured from disentangled LSD profiles is about 2 kG. 

W\,601 was originally classified as a Herbig Be star by \cite{2008A&A...481L..99A} due to its H$\alpha$ emission and mid-infrared excess. However, close examination of H$\alpha$ shows that it varies coherently with \bz, and displays the typical features of a centrifugal magnetosphere. W\,601 is therefore not a classical Herbig Be star. However, both components have low surface gravities, around 3.8. Such a low $\log{g}$ indicates either that the two stars are very evolved (around 22 Myr old) or still contracting towards the main sequence. Only the latter hypothesis is consistent with W\,601's membership in the NGC\,6611 open cluster, which has a main-sequence turnoff age of 2 to 4 Myr reflecting the oldest populations in the cluster, but also contains ongoing star formation. Even if W\,601 is not a member of the cluster, if the stars are on the MS the low $\log{g}$ implies that W\,601\,A must be near the TAMS; given the mass ratio, W\,601\,B should still be near the ZAMS and have a much higher surface gravity than is consistent with the data. Spectroscopic modelling furthermore indicates that the two stars have similar radii, which again is only possible if they are still on the PMS. Therefore, while W\,601 is not a classical Herbig Be star, the orbital and spectoscopic properties of the system indicate that it is probably a PMS star as suggested by \cite{2008A&A...489..459M} and \cite{2008A&A...481L..99A}. 

A rotational period of about 1.13~d was determined from H$\alpha$ and \bz. Modelling of \bz~using this period and W\,601\,B's stellar parameters yields $B_{\rm d} \sim 11$ kG. This makes W\,601\,B by far the most strongly magnetic of the known magnetic PMS stars. When the magnetic star's emission strength is corrected for dilution by the non-magnetic primary, W\,601\,B is also revealed to have H$\alpha$ emission comparable to the strongest seen in the population of CM host stars. This extremely strong emission is consistent with the star's rapid rotation, strong surface magnetic field, mass, and radius. Indeed, the star's emission strength is very close to the value predicted both by empirical \citep{2020MNRAS.499.5379S} and theoretical \citep{2020MNRAS.499.5366O} analyses of similar stars based on the centrifugal breakout mass balancing mechanism.

While \bz~is consistent with a tilted dipole, Stokes $V$ is not satisfactorily reproduced by a purely dipolar model. Strong red-blue asymmetry in the H$\alpha$ emission is furthermore indicative that the surface magnetic field is not purely dipolar. This indicates that distortions to the magnetic dipole are already present on the PMS. 

W\,601\,B is a He-strong star with weakly variable He lines, indicating that surface chemical abundance peculiarities are also already established on the PMS. We were unable to detect indications of significant departure from solar abundances in other spectroscopic lines, or statistically significant variability in the stronger metallic lines. This may indicate that the star is still in the process of forming its surface chemical abundance spots, and that so far only the He spots have become noticeable. On the other hand, the low $S/N$ of the data may simply make metallic abundance patches difficult to detect. 

Extrapolation of W\,601\,B's magnetic and rotational properties as it contracts towards the ZAMS indicate that it should arrive on the MS with a surface dipole magnetic field strength of about 20 kG and a rotational period of about 0.6 d. Thus, when W\,601\,B arrives on the MS, it will have amongst the most extreme magnetic and rotational properties of any known magnetic hot star. W\,601\,B is therefore a precursor to objects such as $\sigma$ Ori E, HR\,5907, or HR\,7355. The expected rate of rotational spin-up, about $-0.5$ s/yr, is similar to the rate recently determined by \cite{2019MNRAS.486.5558S} for HD\,142990 and is in principle detectable if the star is monitored over the next ten or twenty years. 

\section*{Acknowledgements}
The authors acknowledge the numerous helpful comments made by the referee, Dr.\ Gregory Herczeg, which have greatly improved the quality of this paper. This work is based on observations obtained at the Canada-France-Hawaii Telescope (CFHT) which is operated by the National Research Council of Canada, the Institut National des Sciences de l'Univers (INSU) of the Centre National de la Recherche Scientifique (CNRS) of France, and the University of Hawaii. The MiMeS collaboration acknowledges financial support from the Programme National de Physique Stellaire (PNPS) of INSU/CNRS. MES acknowledges financial support from the Annie Jump Cannon Fellowship, supported by the University of Delaware and endowed by the Mount Cuba Astronomical Observatory. EA acknowledges financial support from ‘Programme National de Physique Stellaire’ (PNPS) of CNRS/INSU (France). VP acknowledges support from the National Science Foundation under Grant No.\ 1747658. GAW acknowledges support from the Natural Sciences and Engineering Research Council (NSERC) of Canada in the form of a Discovery Grant.

\section*{Data Availability Statement}
Reduced ESPaDOnS spectra are available at the CFHT archive maintained by the CADC at \url{https://www.cadc-ccda.hia-iha.nrc-cnrc.gc.ca/en/}, where they can be found via standard stellar designations. 

\bibliography{bib_dat.bib}{}

\end{document}